\documentclass[aps,preprint,prd,showpacs,showkeys,nofootinbib,tightenlines]{revtex4-1}

\usepackage{graphicx}  
\usepackage{amsmath}
\usepackage{bm}  
\usepackage{ulem}
\usepackage{amssymb}
\usepackage{comment}
\usepackage{lineno}

\newcommand{\bea}{\begin{eqnarray}}
\newcommand{\eea}{\end{eqnarray}}
\newcommand{\beq}{\begin{equation}}
\newcommand{\eeq}{\end{equation}}
\newcommand{\bqa}{\begin{eqnarray}}
\newcommand{\eqa}{\end{eqnarray}}

\begin{document}

\title{
Production of $\bm{X(3872)}$ Accompanied by a  Pion\\
in $\bm{B}$ Meson Decay}

\author{Eric Braaten}
\email{braaten.1@osu.edu}
\affiliation{Department of Physics,
         The Ohio State University, Columbus, OH\ 43210, USA}

\author{Li-Ping He}
\email{he.1011@buckeyemail.osu.edu}
\affiliation{Department of Physics,
         The Ohio State University, Columbus, OH\ 43210, USA}

\author{Kevin Ingles}
\email{ingles.27@buckeyemail.osu.edu}
\affiliation{Department of Physics,
         The Ohio State University, Columbus, OH\ 43210, USA}

\date{\today}

\begin{abstract}
If the $X(3872)$ is a weakly bound charm-meson molecule, 
it can be produced by the creation of $D^{*0} \bar{D}^0$ or $D^{0} \bar{D}^{*0}$ 
at short distances followed by the formation of  the bound state from the charm-meson pairs.
It can also be produced by the creation of $D^{*}  \bar{D}^*$ at short distances
followed by the  rescattering of the charm mesons into $X \pi$.
We use results of a previous isospin analysis of $B$ meson decays into $K D^{(*)} \bar D^{(*)}$
to estimate the short-distance amplitudes for creating $D^* \bar D^*$.
We use an effective field theory for charm mesons and pions called XEFT to calculate 
the amplitudes for rescattering of  $D^{*}  \bar{D}^*$ into $X \pi$ with small relative momentum.
The  $X\pi$ invariant mass distribution is predicted to have a narrow peak near the $D^{*}  \bar{D}^*$ threshold
from a charm-meson triangle singularity.
We estimate the branching fractions into the peak from the triangle singularity
for the decays $B^0 \to K^+  X \pi^-$ and $B^+ \to K^0  X \pi^+$.
\end{abstract}

\smallskip
\maketitle

\section{Introduction}
\label{sec:Introduction}

The discovery of a large number of exotic hadrons containing a heavy quark and its antiquark 
presents a major challenge to our understanding of QCD
\cite{Chen:2016qju,Hosaka:2016pey,Lebed:2016hpi,Esposito:2016noz,Guo:2017jvc,Ali:2017jda,Olsen:2017bmm,Karliner:2017qhf,Yuan:2018inv,Brambilla:2019esw}.
The $X(3872)$ meson was the first of these exotic hadrons to be discovered.
It is the one for which the most data is available, but there is  still no consensus on its nature. 
The $X$ was discovered in 2003 by the Belle collaboration
in exclusive decays of $B^\pm$ mesons into $K^\pm X$ 
through its decay into $J/\psi\, \pi^+\pi^-$ \cite{Choi:2003ue}.
The observation of its decay into $J/\psi\, \pi^+\pi^-\pi^0$ revealed a dramatic violation of isospin symmetry \cite{Abe:2005ix}.
The $X$ has also been observed in the decay modes  $D^0 \bar D^0 \pi^0$, $D^0 \bar D^0 \gamma$,
$J/\psi\, \gamma$, $\psi(2S)\, \gamma$, and $\chi_{c1} \pi^0$.
The $J^{PC}$ quantum numbers of $X$ were eventually determined to be $1^{++}$ \cite{Aaij:2013zoa}.
Its mass  is extremely close to the $D^{*0} \bar D^0$  threshold,
with the difference being only $0.01 \pm 0.18$~MeV \cite{Tanabashi:2018oca}.
This suggests that $X$ is a weakly bound S-wave charm-meson molecule with the flavor structure
\begin{equation}
\big| X(3872) \big\rangle = 
\frac{1}{\sqrt2} \Big( \big|D^{*0} \bar D^0 \big\rangle + \big|D^0 \bar D^{*0}\big\rangle \Big).
\label{X-flavor}
\end{equation}
There are alternative models for the $X$
\cite{Chen:2016qju,Hosaka:2016pey,Lebed:2016hpi,Esposito:2016noz,Guo:2017jvc,Ali:2017jda,Olsen:2017bmm,Karliner:2017qhf,Yuan:2018inv,Brambilla:2019esw},
but the observation of $X$ in 7 different decay modes has not been effective in discriminating between these models.
However there may be aspects of the production of $X$ that are more effective at discriminating
between models than the decays of $X$.

A convenient theoretical framework for describing  $X$ as a weakly bound charm-meson molecule
is an effective field theory for charm mesons and pions called XEFT  \cite{Fleming:2007rp}.
It describes the sector of QCD consisting of
$D^* \bar D$, $D \bar D^*$,  and $D \bar D \pi$ with small relative momenta as well as the weakly bound state $X$.
In Ref.~\cite{Braaten:2010mg}, it was pointed out that XEFT 
could also be applied to the sector of QCD consisting of
$D^* \bar D^*$, $D \bar D^* \pi$, $D^* \bar D \pi$, $D \bar D \pi \pi$, and $X\pi$.
XEFT can be applied to the production of $X$ from short-distance processes that create a pair of charm mesons.
If a high energy reaction creates $D^{*0} \bar D^0$ and $D^0 \bar D^{*0}$ at short distances,
XEFT can describe their binding into  $X$.
If a high energy reaction creates   $D^* \bar D^*$ at short distances,  XEFT can describe their rescattering into $X\pi$.

In Ref.~\cite{Braaten:2019sxh}, we pointed out that the prompt production of $X$ 
accompanied by a  pion could be important at a hadron collider.  
We calculated the  cross sections
for inclusive production of $X \pi^\pm$ and $X \pi^0$ with small relative momentum.
The calculations took advantage of cancellations of interference effects
from the sum over the many additional particles in the inclusive cross sections.
The $X \pi$ invariant mass distribution has a narrow peak near  the $D^* \bar D^*$ threshold. 
In retrospect, these narrow peaks come from 
triangle singularities \cite{Szczepaniak:2015eza,Liu:2015taa,Szczepaniak:2015hya,Guo:2017wzr}.
The corresponding Feynman diagrams have three charm meson lines that form a triangle,
and this results in a kinematic singularity from the region where all three charm mesons are on-shell.

Guo recently pointed out that if a short-distance process can create an S-wave $D^* \bar D^*$ pair,
it will produce a narrow peak in the $X \gamma$ invariant mass distribution near the $D^* \bar D^*$ threshold
from a charm-meson triangle singularity \cite{Guo:2019qcn}.
Dubinskiy and Voloshin pointed out previously that $e^+ e^-$ annihilation will produce a narrow peak in the $X \gamma$ 
invariant mass distribution from rescattering of a P-wave $D^{*0} \bar D^{*0}$ pair \cite{Dubynskiy:2006cj}.
The peak comes from a charm-meson triangle singularity.
In Ref.~\cite{Braaten:2019gfj}, we predicted the normalized cross section near the peak,
and we showed that the peak may be large enough to be observed by the BESIII detector.

Another short-distance process that can create a $D^* \bar D^*$ pair is $B$ meson decay. 
In this paper, we study exclusive decays of $B$ mesons  into $KX\pi$ through the decay into $KD^* \bar D^*$
at short distances followed by the rescattering of  $D^* \bar D^*$  into $X\pi$.
In Section~\ref{sec:XEFT}, we  summarize previous work on the effective field theory XEFT. 
In Section~\ref{sec:BtoKDD}, we describe
a precise  isospin analysis of the decays $B \to K D^{(*)} \bar D^{(*)}$ by Poireau and Zito \cite{Poireau:2011gv}.
In Section~\ref{sec:BtoKX}, we verify that measurements for $B \to KX$ are compatible 
with the isospin amplitudes of Poireau and Zito for decays into $K D^{*0} \bar D^0$ and $K D^{0} \bar D^{*0}$. 
In Section~\ref{sec:HQS}, we construct interaction terms for $B \to K D^{(*)} \bar D^{(*)}$ 
in which the $c \bar c$ pair in the charm mesons are in a spin-triplet state
when the relative momentum of the charm mesons is 0.
In Section~\ref{sec:BtoKD*D*0}, we use XEFT to calculate the rates for producing $D^* \bar D^*$ near the threshold.
In Section~\ref{sec:BtoKXpi}, we use XEFT to calculate the rates for the rescattering of  $D^* \bar D^*$  into $X\pi$.
We conclude  in Section~\ref{sec:Conclusion}  with a discussion of our results.


\section{XEFT}
\label{sec:XEFT}

The difference between the mass of the $X(3872)$ 
and the energy of the $D^{*0} \bar D^0$ scattering threshold is \cite{Tanabashi:2018oca}
\begin{equation}
E_X \equiv M_X - (M_{*0}+M_0) =( +0.01 \pm 0.18)~\mathrm{MeV}.
\label{eq:deltaMX}
\end{equation}
We denote the masses of the spin-0 charm mesons  $D^0$ and $D^+$ by $M_0$ and $M_1$
(or collectively by $M_{D}$),
the masses of the spin-1 charm mesons  $D^{*0}$ and $D^{*+}$ by $M_{*0}$ and $M_{*1}$
(or collectively by $M_{D^*}$),
and the masses of the pions  $\pi^0$ and $\pi^+$ by $m_0$ and $m_1$
(or collectively by $m_\pi$).
The reduced mass of $D^{*0}$ and $\bar D^0$ is $\mu=M_{*0}M_0/(M_{*0}+M_0)$.
The central value in Eq.~\eqref{eq:deltaMX} corresponds to on-shell charm mesons,
which would require the $X$ to be a virtual state rather than a bound state.
The value lower by $1\sigma$ corresponds to a bound state with binding energy  $|E_X|=0.17$~MeV
and binding momentum  $\gamma_X \equiv \sqrt{2 \mu |E_X|} =18$~MeV. 

If short-range interactions produce an S-wave bound state very close to the scattering threshold for its constituents,
the few-body physics has universal aspects determined by the binding momentum $\gamma_X$ \cite{Braaten:2004rn}.
The  universal  wavefunction for the constituents of the bound state to have 
 relative  momentum $k$ small compared to the inverse range is
\begin{equation}
\psi_X(k) = \frac{ \sqrt{8 \pi \gamma_X}}{k^2 + \gamma_X^2}.
\label{psiX-k}
\end{equation}
The universal scattering amplitude for the elastic scattering of the constituents with
relative momentum $k$ small compared to the inverse range is
\begin{equation}
f_X(k) = \frac{1}{-\gamma_X- i k}.
\label{f-E}
\end{equation}

The universal results in Eqs.~\eqref{psiX-k} and  \eqref{f-E}
can be derived from a {\it zero-range effective field theory} \cite{Braaten:2005jj}. 
In the case of the $X$, it is a nonrelativistic effective field theory (EFT) for the 
neutral charm mesons $D^{*0}$, $\bar D^{*0}$, $D^0$, and $\bar D^0$.
This EFT describes explicitly the  $D^{*0} \bar D^0$ and  $D^{0} \bar D^{*0}$ 
components of the $X$.  Since the EFT does not describe charged charm mesons explicitly,
its range of validity extends  in energy at most up to the $D^{*+} D^-$ scattering threshold,
which is higher than the $D^{*0}\bar D^0$ scattering threshold by 8.2 MeV. 
This EFT  does not describe explicitly the $D^{0} \bar D^0 \pi^0$ component of the $X$, 
which can arise from the decays $D^{*0} \to D^0 \pi^0$ or $\bar D^{*0} \to \bar D^0 \pi^0$. 

Fleming, Kusunoki, Mehen and van Kolck developed a nonrelativistic effective field theory 
called {\it XEFT} with a much greater range of validity than the zero-range EFT,
because it describes pion interactions explicitly \cite{Fleming:2007rp}.
It is an EFT for neutral and charged S-wave charm mesons 
$D^*$, $\bar D^*$, $D$, and $\bar D$ and for neutral and charged pions $\pi$.
The number of charm mesons $D$ and $D^*$ and the number of anti-charm mesons  $\bar D$ and $\bar D^*$
are conserved in XEFT.
The contact interactions among the charm-meson pairs $D^* \bar D$  and $D \bar D^*$
in the $J^{PC} = 1^{++}$ channel with total electric charge 0 must be treated nonperturbatively in XEFT,
but the coupling constant for pion interactions is small enough that the transitions $D^*\leftrightarrow D \pi$
can be treated perturbatively  \cite{Fleming:2007rp}.
XEFT describes explicitly the  $D^* \bar D$, $D \bar D^{*}$, and $\bar D D \pi$ components of the $X$.
If a high energy process creates $D^{*0} \bar D^0$ and $D^0 \bar D^{*0}$ at short distances,
XEFT can describe the subsequent formation of $X$ by the binding of the charm mesons.
The region of validity of the original formulation of XEFT extends  up to about the minimum energy 
required to produce  a $\rho$ meson.
For a charm meson pair, this corresponds to a relative momentum  greater than 1000~MeV. 
For a charm meson pair plus a pion, the region of validity of XEFT is also  limited by the nonrelativistic approximation
for the pion: the relative momentum of the pion must be less than about $m_\pi \approx 140$~MeV.
We refer to a  pion with relative momentum  of order $m_\pi$ or smaller as a {\it soft pion}.
 
Although pion interactions can be treated perturbatively in XEFT, they can also be treated nonperturbatively.
The $D \bar D \pi$ components of the $X$  have been taken into account with nonperturbative pion interactions 
by solving Faddeev integral equations \cite{Baru:2011rs}.
The intensively numerical character of this approach makes it difficult to extract  simple physical predictions.

A Galilean-invariant formulation of XEFT that exploits the 
approximate conservation of mass in the transitions $D^* \leftrightarrow D \pi$
was developed in Ref.~\cite{Braaten:2015tga}.
In  Galilean-invariant XEFT, the spin-0 charm mesons $D^0$ and $D^+$ have the same kinetic mass $M_0$,
the spin-1 charm mesons $D^{*0}$ and $D^{+}$ have the same kinetic mass $M_0+m_0$,
and the pions $\pi^0$ and $\pi^+$  have the same kinetic mass $m_0$.
The difference between the physical mass and the kinetic mass of a particle is taken into account through its rest energy.
The pion number defined by the sum of the numbers of $D^*$, $\bar D^*$, and $\pi$ mesons
is conserved in Galilean-invariant XEFT.
The region of validity of Galilean-invariant XEFT extends  up to about the minimum energy required 
to produce an additional pion, which is above the  $D^* \bar D$ threshold by about 140~MeV. 
Galilean invariance also simplifies the utraviolet divergences of XEFT.

An alternative Galilean-invariant EFT for S-wave charm mesons and pions that may be more predictive 
has been introduced by Schmidt, Jansen, and Hammer \cite{Schmidt:2018vvl}.
The only fields in this EFT are those for the spin-0 charm mesons $D$ and the pions $\pi$.
The spin-1 charm mesons $D^*$ arise dynamically as P-wave $D \pi$ resonances.

In Ref.~\cite{Braaten:2010mg}, Braaten, Hammer, and Mehen pointed out that XEFT 
could also be applied to sectors with pion number larger than 1.
In particular, it can be applied to the sector with pion number 2,
which consists of $D^* \bar D^*$, $D \bar D^* \pi$, $D^* \bar D \pi$, $D \bar D \pi \pi$, and $X \pi$.
The cross sections for $D^* \bar D^* \to D^* \bar D^*$ and $D^* \bar D^* \to X \pi$ at small kinetic energies
were calculated in Ref.~\cite{Braaten:2010mg}.
If a high energy process can create $D^* \bar D^*$  at short distances,
XEFT can describe their subsequent rescattering into $X$ plus a soft pion.
The inclusive prompt production of  $X$ plus a soft pion in high-energy hadron collisions was discussed 
 in Ref.~\cite{Braaten:2019sxh}.
In this paper, we consider the production of $X$ plus a soft pion in the exclusive decay of a $B$ meson into $KX\pi$.


\section{Decays into  $\bm{K}$ plus a Charm-Meson Pair}
\label{sec:BtoKDD}

A $B$ meson can decay into a kaon and a pair of charm mesons.
The symmetries of QCD provide constraints on the matrix elements for the decays $B \to K D^{(*)} \bar D^{(*)}$.  
The only exact symmetry is Lorentz invariance, which requires a matrix element to be a Lorentz-scalar
function of the 4-momenta $k$, $p$, and $\bar p$ of $K$, $D^{(*)}$, and $ \bar D^{(*)}$
and the polarization 4-vectors $\varepsilon$ and $\bar \varepsilon$ of $D^*$ and $\bar D^*$.
If the square of the matrix element is summed over the spins of any spin-1 charm mesons $D^*$ or $\bar D^*$,
it reduces to a function of the invariant masses $(p+\bar p)^2$ of $D^{(*)} \bar D^{(*)}$ and $(k+p)^2$ of $K D^{(*)}$.
The graphical representation of the dependence on these two variables is called a {\it Dalitz plot}.

The approximate isospin symmetry of QCD provides strong constraints on  the matrix elements
for the decays $B \to K D^{(*)} \bar D^{(*)}$.  Each of the particles in such a reaction is a member of 
an isospin doublet. At the quark level, the decays for $B^+$ and $B^0$
are $\bar b q_1 \to (\bar s q_2)(c \bar q_3) (\bar c q_4)$, where each $q_i$ is $u$ or $d$.
The isospin doublets for the light quarks and antiquarks are
\begin{equation}
\binom{u}{d}, \quad \binom{-\bar d}{\bar u}.
\end{equation}
The isospin doublets for the $B$ meson, the kaon, and the spin-0 charm mesons $D$ and $\bar D$ are
\begin{equation}
\binom{B^+}{B^0}, \quad \binom{K^+}{K^0}, \quad \binom{-D^+}{D^0}, \quad \binom{\bar D^0}{D^-}.
\end{equation}
The isospin doublets for the spin-1 charm mesons $D^*$ and $\bar D^*$ are analogous
to those for $D$ and $\bar D$.
The $SU(2)$ isospin symmetry reduces the  matrix elements to two complex amplitudes 
for each of the 4 sets of channels $K D \bar D$, $K D^* \bar D$, $K D\bar D^*$, and $K D^* \bar D^*$. 
One  choice for the isospin amplitudes $A_0$ and $A_1$ corresponds to $D^{(*)}K$ in  
an isospin-singlet and isospin-triplet state, respectively.

The expressions for the decay rates for $B \to \, KD^{(*)} \bar D^{(*)}$
in terms of dimensionless  Lorentz-invariant matrix elements $\mathcal{A}$ are
\begin{equation}
\Gamma\big[B \to  K  D^{(*)} \bar D^{(*)}\big] = 
\frac{1}{2 M_B}\int \!d\Phi_{K  D^{(*)} \bar D^{(*)}}\,
\Big|\mathcal{A}\big[B \to  K  D^{(*)} \bar D^{(*)}\big]\Big|^2,
\label{GammaBtoKDD}
\end{equation}
where $M_B$ is the mass of the $B$ meson and 
$d\Phi_{K  D^{(*)} \bar D^{(*)}}$ is the differential phase space for the three mesons in the final state.
Factors of 3 from summing over spins of $D^*$ or $\bar D^*$ are absorbed into the amplitudes $\mathcal{A}$.
Using isospin symmetry, the amplitudes for the decays of $B$ into  $K D^{*0} \bar D^0$ and $KD^0 \bar D^{*0}$
can be expressed in terms of 4 complex isospin amplitudes   \cite{Zito:2004kz}:
\begin{subequations}
\begin{eqnarray}
\mathcal{A}\big[B^0 \to  K^0  D^{*0} \bar D^0\big] &=& 
- \sqrt{\tfrac23} \, A_1^{L*},
\\
\mathcal{A}\big[B^0 \to  K^0  D^0 \bar D^{*0}\big] &=& 
- \sqrt{\tfrac23} \, A_1^{*L} ,
\\
\mathcal{A}\big[B^+ \to K^+ D^{*0} \bar D^0 \big] &=& 
\sqrt{\tfrac16} \, A_1^{L*} + \sqrt{\tfrac12} \, A_0^{L*},
\label{APZ:K+D*0D0}
\\
\mathcal{A}\big[B^+ \to K^+ D^0 \bar D^{*0} \big] &=& 
\sqrt{\tfrac16} \, A_1^{*L} + \sqrt{\tfrac12} \, A_0^{*L}.
\end{eqnarray}
\label{ABKD0D0}%
\end{subequations}
These four amplitudes will be applied to the decays $B \to K X(3872)$ in Section~\ref{sec:BtoKX}.
The amplitudes for the decays $B \to K D^{*} \bar D^*$ can be expressed 
 in terms of 2 complex isospin amplitudes \cite{Zito:2004kz}:
\begin{subequations}
\begin{eqnarray}
\mathcal{A}\big[B^0 \to  K^0 D^{*0} \bar D^{*0}\big] &=& 
- \sqrt{\tfrac23}\, A_1^{**} ,
\label{AB0K0D*0D*0}
\\
\mathcal{A}\big[B^0 \to  K^0 D^{*+} D^{*-}\big] &=& 
\sqrt{\tfrac16}\,  A_1^{**}+ \sqrt{\tfrac12}\,  A_0^{**} ,
\\
\mathcal{A}\big[B^0 \to  K^+  D^{*0} D^{*-}\big] &=& 
\sqrt{\tfrac16}\,  A_1^{**}- \sqrt{\tfrac12}\,  A_0^{**} ,
\label{AB0K+D*0D*-}
\\
\mathcal{A}\big[B^+ \to K^+  D^{*0} \bar D^{*0}\big] &=& 
\sqrt{\tfrac16}\,  A_1^{**} +\sqrt{\tfrac12}\,  A_0^{**} ,
\label{ABpK+D*0D*0}
\\
\mathcal{A}\big[B^+ \to K^+   D^{*+} D^{*-}\big] &=& 
- \sqrt{\tfrac23}\, A_1^{**} ,
\\
\mathcal{A}\big[B^+ \to K^0  D^{*+} \bar D^{*0}\big] &=& 
\sqrt{\tfrac16}\,  A_1^{**}  - \sqrt{\tfrac12}\,  A_0^{**} .
\end{eqnarray}
\label{ABKD*D*}%
\end{subequations}
The four  amplitudes for the decays into final states that include $D^{*0}$ or $\bar D^{*0}$ will
be applied to the decays  $B \to \, KX \pi$  in Section~\ref{sec:BtoKXpi}.

\begin{table}[t]
\centering
\begin{tabular}{l|c c c}
            &  $|A_0| \times 10^5$   &  $|A_1| \times 10^5$  &     $\delta$      \\
            \hline
$L*$~   &   ~$1.33 \pm 0.04$~   &  ~$0.42 \pm 0.04$~   &  ~$0.925 \pm 0.157$~   \\
$*L$     &     $0.92 \pm 0.03$     &     $0.41 \pm 0.04$     &    $1.798 \pm 0.122$\\
$**$     &     $2.28 \pm 0.08$      &     $0.72 \pm 0.05$     &    $1.745 \pm 0.122$ \\
\end{tabular}
\caption{Amplitudes for $B \to K \bar{D}^{(*)} D^{(*)}$ decays from Ref.~\cite{Poireau:2011gv}. 
The rows labeled $L*$, $*L$, and $**$ correspond to the $D^* \bar D$, $D \bar D^*$,
and $D^* \bar D^*$ channels, respectively. The  complex phase $e^{i \delta}$ of 
$A_1/A_0$ defines the  angle $\delta$.}
\label{tab:PZamps}
\end{table}

If the squares of the amplitudes  in Eq.~\eqref{GammaBtoKDD} are 
summed over the spin states of any spin-1 charm meson $D^* $ or $\bar D^*$
and averaged over the Dalitz plot, the corresponding branching fractions reduce to
\begin{equation}
\mathrm{Br}\big[B \to  K  D^{(*)} \bar D^{(*)}\big] =
\frac{\tau[B]}{2 M_B}
\Big|\mathcal{A}\big[B \to  K  D^{(*)} \bar D^{(*)}\big]\Big|^2\,  \Phi_{K  D^{(*)} \bar D^{(*)}},
\label{BrBtoKDD}
\end{equation}
where $\tau[B]$ is the lifetime of the $B$ meson
and $\Phi_{K  D^{(*)} \bar D^{(*)}}$ is the integrated 3-body phase space. 
The ratio of the $B^+$ and $B^0$ lifetimes is 
$\tau[B^+]/\tau[B^0] = 1.076 \pm 0.004$ \cite{Tanabashi:2018oca}.

A precise isospin analysis of the decays $B\to K D^{(*)}\bar D^{(*)}$
has been presented by Poireau and Zito  \cite{Poireau:2011gv}.
The analysis used measurements of 22 branching fractions
by the BaBar collaboration \cite{delAmoSanchez:2010pg}
and measurements of 2 branching fractions by the Belle collaboration \cite{Dalseno:2007hx,Brodzicka:2007aa}.
For each of the four sets of decay channels $KD \bar D$, $KD^* \bar D$, $KD \bar D^*$, and $KD^* \bar D^*$,
Poireau and Zito determined the absolute values and the relative phase 
of two complex isospin amplitudes $A_0$ and $A_1$  by fitting the expressions for the branching fractions 
in Eqs.~\eqref{BrBtoKDD} to the measurements by the BaBar and Belle collaborations.
The isospin amplitudes that appear in Eqs.~\eqref{ABKD0D0} and \eqref{ABKD*D*} are given in Table~\ref{tab:PZamps}.

The separation of scales in the matrix elements for decays $B \to K D^{(*)} \bar D^{(*)}$ 
would allow them to be expressed as products of short-distance factors involving momenta
of order $m_\pi$ or larger and long-distance factors involving only smaller momentum scales.
Summing  the squares of matrix elements over the spin states of any $D^* $ or $\bar D^*$
and then averaging them over the Dalitz plot, as in the analysis of Ref.~\cite{Poireau:2011gv},
 decreases their sensitivity to long-distance effects, such as resonances.
We  will use the constant amplitudes of  Poireau and Zito  as approximations to
short-distance amplitudes for the decays $B \to KD^{(*)} \bar D^{(*)}$ in the region of the Dalitz plot 
where the charm-meson pair has small relative momentum.


\section{Decays into $\bm{K}$ plus $\bm{X}$}
\label{sec:BtoKX}

The flavor structure of the $X(3872)$ in Eq.~\eqref{X-flavor}  implies that
the amplitude for producing $X$  is proportional  to the sum of the complex amplitudes for 
producing $D^{*0} \bar D^0$ and $D^0 \bar D^{*0}$. 
In the decay of a $B$ meson into $K D^{*0} \bar D^0$ or $K D^0 \bar D^{*0}$ with the charm-meson pair 
having small relative momentum, the momentum in the charm-meson-pair rest frame 
of either the incoming $B$ or the  outgoing $K$  is  about 1550~MeV.
Since this momentum is much larger than the pion mass $m_\pi \approx 140$~MeV,
the $B$-to-$K$ transition that creates the charm mesons occurs over distances much shorter than the range $1/m_\pi$
of the interactions between the charm mesons.
The interactions between $D^{*0} \bar D^0$ and between $D^0 \bar D^{*0}$ also
involve the scale $\gamma_X$ of the binding momentum of the $X$, which is much smaller than $m_\pi$.  
The amplitude for the decay can therefore be factored into a long-distance factor that involves
$\gamma_X$ and a  short-distance factor that involves only momentum scales of order $m_\pi$ or larger.
In Ref.~\cite{Artoisenet:2009wk}, the inclusive prompt cross sections in high energy hadron collisions for producing 
$D^{*0} \bar D^0$ with small relative momentum and for producing $X$
were expressed in factored forms, with long-distance factors that involve  $\gamma_X$ 
and short-distance factors that involve only momentum scales of order $m_\pi$ or larger.
The analogous  factored form for the exclusive decay rate of $B$ into  $KX$ is
\begin{equation}
\Gamma \big[B \to K X \big] =
\frac{1}{2 M_B} \int d\Phi_{(D^*\bar D)K}\, 
\left| \frac{\mathcal{A} \big[KD^{*0} \bar D^0 \big] + \mathcal{A} \big[KD^0 \bar D^{*0} \big]}{\sqrt2}\right|^2 
\frac{\Lambda^2\gamma_X}{4\pi\mu},
\label{factor-X}
\end{equation}
where  $\mu$ is the reduced mass of $D^{*0}$ and $\bar D^0$
and $d\Phi_{(D^*\bar D)K}$ is the  differential two-body phase space 
for $K$ and a composite particle denoted by  $(D^*\bar D)$ with mass $M_{D^*} \!+\! M_D$.
Factors of 3 from the sums over the spin states of $D^{*0}$ or $\bar D^{*0}$
are absorbed into the amplitudes $ \mathcal{A}$.
The short-distance factor in Eq.~\eqref{factor-X} 
involves the short-distance amplitudes $\mathcal{A}[KD^{*0} \bar D^0]$ 
and $ \mathcal{A}[KD^0 \bar D^{*0}]$ for producing $D^{*0} \bar D^0$ and $D^0 \bar D^{*0}$.
The short-distance factor also   includes the square of an unknown momentum scale $\Lambda$ of order $m_\pi$.
The factor $\Lambda^2$ is not universal.  The corresponding factor 
in another short-distance production rate may have a different value of order $m_\pi^2$. 
In the case of inclusive prompt production of $X$ at high-energy hadron colliders,
the sums over the many additional particles in the final state wash out the interference
between the amplitudes for producing $D^{*0} \bar D^0$ and $D^0 \bar D^{*0}$  
and make their contributions to the cross section  approximately equal.
In the case of exclusive decays of the $B$ meson, the interference effects can be important.

The short-distance amplitudes $\mathcal{A}[KD^{*0} \bar D^0]$ and $\mathcal{A}[KD^0 \bar D^{*0}]$
in Eq.~\eqref{factor-X} can be expressed in terms of isospin amplitudes as in Eqs.~\eqref{ABKD0D0}.
The resulting expressions for the  decay rates for $B^+ \to K^+  X$ and $B^0 \to K^0 X$ are
\begin{subequations}
\begin{eqnarray}
\Gamma\big[B^+ \to  K^+  X\big]
&=&  \frac{ \lambda^{1/2}(M_B , M_{*0} \!+\! M_0,m_K )\, \Lambda^2 \gamma_X}{768\pi^2 M_B^3 \mu} 
\big| A_1^{L*}+ A_1^{*L}  + \sqrt{3} \, (A_0^{L*} +  A_0^{*L}) \big|^2,
 \label{GammaBptoKpX}
\\
\Gamma\big[B^0 \to  K^0 X\big]
&=&  \frac{\lambda^{1/2}(M_B , M_{*0} \!+\! M_0,m_K )\, \Lambda^2 \gamma_X}{192\pi^2 M_B^3 \mu}
 \, \big| A_1^{L*} + A_1^{*L} \big|^2,
 \label{GammaB0toK0X}
\end{eqnarray}
 \label{GammaBtoKX}%
\end{subequations}
 where $\lambda (x,y,z) = (x^4+y^4+z^4)-2(x^2y^2+y^2z^2+z^2x^2)$.
The ratio of  the branching fractions for these decays reduces to
\begin{eqnarray}
\frac{\mathrm{Br}\big[B^+ \to  K^+  X\big]}{\mathrm{Br}\big[B^0 \to  K^0 X\big]}
&=& \frac{\tau[B^+]}{\tau[B^0]}\,
\frac{\big| A_1^{L*}+ A_1^{*L}  + \sqrt{3} \, (A_0^{L*} +  A_0^{*L}) \big|^2}
       {4 \big| A_1^{L*} + A_1^{*L} \big|^2}.
 \label{BRBKX}
\end{eqnarray}
An experimental result for the branching ratio in Eq.~\eqref{BRBKX}
can be obtained from measurements of the products of the branching fractions for $B \to  K  X$
and the branching fraction for $X \to J/\psi \, \pi^+\pi^-$ \cite{Tanabashi:2018oca}:
\begin{equation}
\frac{\mathrm{Br}\big[B^+ \to  K^+  X\big]}{\mathrm{Br}\big[B^0 \to  K^0 X\big]}
= 2.00 \pm 0.63.
 \label{BRBKX:exp}
\end{equation}

\begin{figure}[t]
\includegraphics*[width=0.8\linewidth]{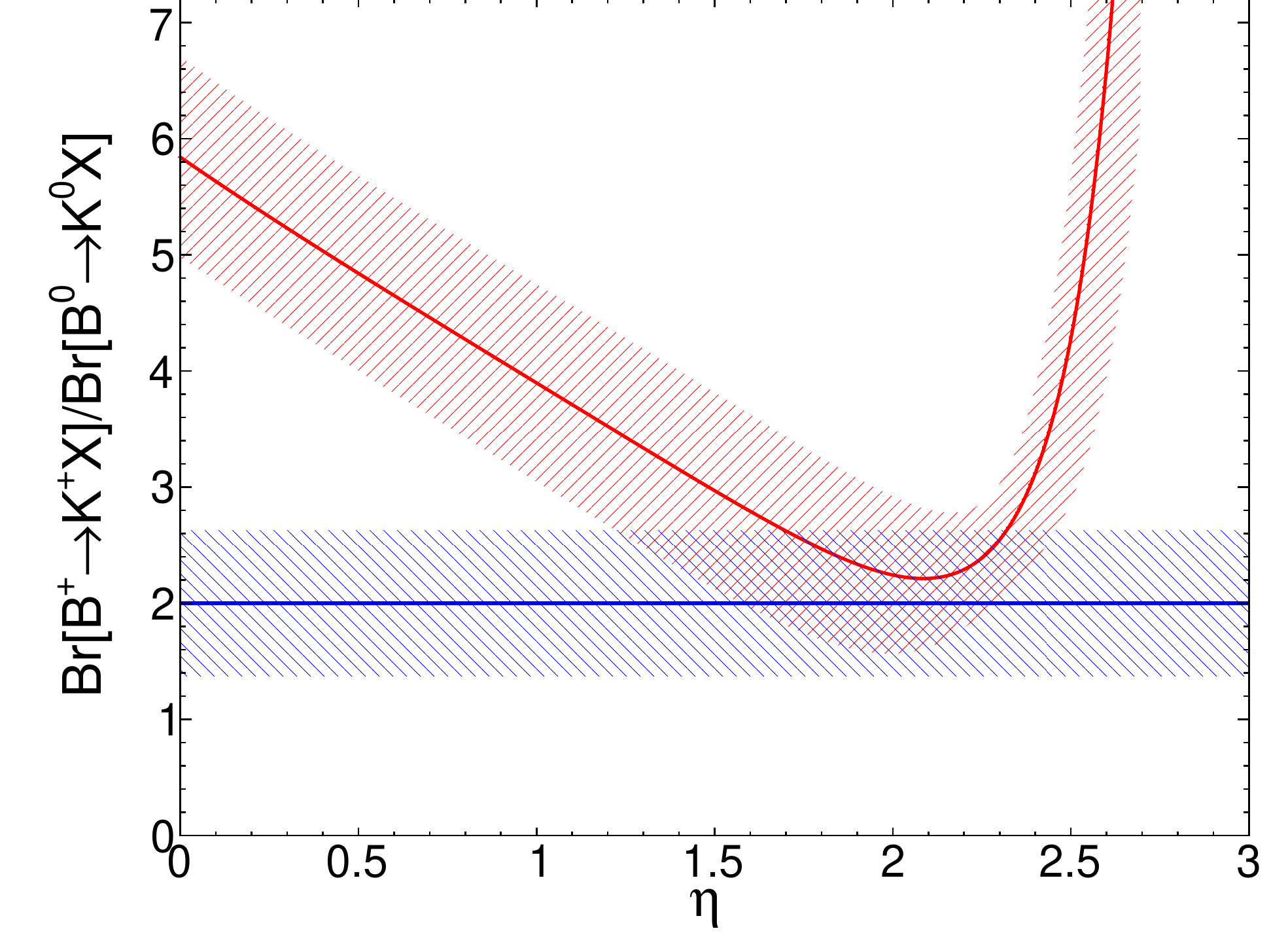} 
\caption{
Ratio of the branching fractions for $B^+ \to  K^+  X$ and $B^0 \to  K^0 X$
as a function of the angle $\eta$ in the complex phase of $A_1^{L*}/A_1^{*L}$.
The solid red curve is the theoretical prediction using the central values of the amplitudes 
in Table~\ref{tab:PZamps}, and the hatched region is the associated error band. 
The horizontal band is the experimental result in Eq.~\eqref{BRBKX:exp}.
}
\label{fig:BRp0-eta}
\end{figure}

The theoretical result for the ratio of  branching fractions in Eq.~\eqref{BRBKX}
depends on the short-distance isospin amplitudes $A_0^{L*}$, $A_1^{L*}$, $A_0^{*L}$, and $A_1^{*L}$.
We will approximate these short-distance isospin amplitudes  by
the isospin amplitudes determined by the analysis in  Ref.~\cite{Poireau:2011gv}.
The absolute values of these isospin amplitudes and  the complex phases 
of $A^{L*}_1/A^{L*}_0$ and $A^{L*}_1/A^{L*}_0$ are given with error bars in Table~\ref{tab:PZamps}.
The ratio also depends on the complex phase $e^{i \eta}$ of  $A_1^{L*}/A_1^{*L}$,
which was not determined in Ref.~\cite{Poireau:2011gv}.
We assume for simplicity that all the error bars in Table~\ref{tab:PZamps}
and in the ratio  $\tau[B^+]/\tau[B^0]$ are uncorrelated Gaussian errors.
The ratio of  branching fractions in Eq.~\eqref{BRBKX} can then be predicted as a function of $\eta$ with
errors  by combining all the errors in quadrature. 
The  theoretical prediction is close to the experimental result  in Eq.~\eqref{BRBKX:exp}
only if the angle $\eta$ in the phase factor $e^{i \eta}$ is close to 2.
 In Fig.~\ref{fig:BRp0-eta}, the theoretical prediction 
 is shown as a function of $\eta$  in the region near $\eta = 2$ along with the experimental error band.
The difference between the central values in Fig.~\ref{fig:BRp0-eta} is 
the fewest number of standard deviations at $\eta = 2.07$, where the difference is $0.24\, \sigma$. 
By requiring the difference between the theoretical prediction and
the experimental result to be less than $1\, \sigma$, we obtain  error bars on the angle $\eta$:
\begin{equation}
\eta = 2.07^{+0.30}_{-0.62}.
 \label{eta}
\end{equation}

Having determined the angle $\eta$ in Eq.~\eqref{eta}, we can quantify the effects of interference 
in the decay rates for $B$ mesons into $K X$ in  Eq.~\eqref{factor-X}.
For $B^0$ decays, the central values of the squares of the absolute values of the amplitudes
$\mathcal{A}[K^0D^{*0} \bar D^0]$, $\mathcal{A}[K^0D^0 \bar D^{*0}]$, and their sum
are  0.118, 0.112, and  0.120 times $10^{-10}$, respectively. 
Since the sum of the first two is approximately twice the third,  there is substantial destructive interference.
For $B^+$ decays, the central values of the squares of the absolute values of the amplitudes
$\mathcal{A}[K^+D^{*0} \bar D^0]$, $\mathcal{A}[K^+D^0 \bar D^{*0}]$, and their sum
are 1.11, 0.40, and 0.25 times $10^{-10}$, respectively. 
Since the sum of the first two is much greater than the third, 
there is large destructive interference.

We proceed to make a quantitative estimate of the branching fraction for $B^0 \to  K^0 X$.
After inserting the central values of the isospin amplitudes and  $\eta$
into the decay rate in Eq.~\eqref{GammaB0toK0X}, the branching fraction is
\begin{equation}
\mathrm{Br}\big[B^0 \to  K^0 X\big]
\approx \big( 6.5 \times 10^{-7} \big) 
\left( \frac{\Lambda}{m_\pi} \right)^2 \left(\frac{|E_X|}{0.17\, \mathrm{MeV}} \right)^{1/2}.
 \label{BrB0toK0X}
\end{equation}
The error in the prefactor from combining the errors in the isospin amplitudes and $\eta$ in quadrature is more than 100\%,
with most of the error coming from $\eta$.
The measured product of this branching fraction with that for the decay of $X$ into $J/\psi\, \pi^+\pi^-$ is 
\cite{Tanabashi:2018oca}
\begin{equation}
\mathrm{Br}\big[B^0 \to  K^0 X\big]\,\mathrm{Br}\big[X \to J/\psi\, \pi^+\pi^-\big]
=   ( 4.3 \pm 1.3 ) \times 10^{-6}.
 \label{BrBrB0toK0X}
\end{equation}
In Ref.~\cite{Braaten:2019ags}, we derived upper and lower bounds on the branching fraction Br 
for the $X$ bound state to decay into $J/\psi\, \pi^+\pi^-$.
The loose lower bound $\mathrm{Br} > 4\%$ can be derived from a recent measurement by the BaBar collaboration
of the inclusive branching fraction for $B^+$ into $K^+$ plus the  $X$ resonance feature \cite{Wormser}. 
An upper bound $\mathrm{Br} < 33\%$ can be derived from measurements of branching ratios 
of $J/\psi\, \pi^+\pi^-$ over other short-distance decay modes of the $X$.
Given the undetermined binding  energy $|E_X|$,
 the large error in the prefactor in Eq.~\eqref{BrB0toK0X},
and the uncertainty in the branching fraction for $X \to J/\psi\, \pi^+\pi^-$,
the best we can say is that the estimate of the branching fraction for $B^0 \to  K^0 X$ in Eq.~\eqref{BrB0toK0X}
 is compatible with the measurement in Eq.~\eqref{BrBrB0toK0X} for some value of $\Lambda$ of order $m_\pi$.


\section{Heavy Quark Symmetries}
\label{sec:HQS}

The isospin analysis of Poireau and Zito in Ref.~\cite{Poireau:2011gv}
exploited the isospin symmetry of QCD.
There are other approximate symmetries of QCD that can be used to constrain 
the matrix elements for the decays $B \to K D^{(*)} \bar D^{(*)}$.  
One of them is the approximate $SU(3)_L \times SU(3)_R$ chiral symmetry.
The $K$ is a pseudo-Goldstone boson associated with the spontaneous breaking of this symmetry,
so a matrix element must vanish in the limit as the 4-momentum of $K$ goes to 0.
This constraint is automatically satisfied if the matrix element has a factor of the 4-momentum $k^\mu$ of $K$.

Heavy-quark symmetries are approximate symmetries of QCD  that  relate matrix elements 
between the 4 sets of channels $K D \bar D$, $K D^* \bar D$, $K D\bar D^*$, and $K D^* \bar D^*$. 
The constraints of heavy-quark symmetries can be expressed most conveniently
by arranging the Lorentz-scalar field $D(x)$ for a $D$ and the Lorentz-vector field $D^\mu(x)$ for a $D^*$
into a charm-meson multiplet field $H^{(c)}(x)$ that is a $4\times 4$ matrix. 
In momentum space, the spin-1 charm-meson field $D^\mu(p)$ with 4-momentum $p$
satisfies the constraint $p_\mu D^\mu = 0$.
The charm-meson multiplet field that creates $D$ or $D^*$ with 4-velocity $v$  and 
the anticharm-meson multiplet field  that creates $\bar D$ or $\bar D^*$  with 4-velocity $\bar v$
are \cite{Grinstein:1992qt}
\begin{subequations}
\begin{eqnarray}
\bar H^{(c)}_c(v)&=&
\big[ D^{\mu}_c(v)^\dagger\,  \gamma_\mu + D_c(v)^\dagger \, \gamma_5 \big]  \frac{1+  v\!\!\!\slash}{2},
\label{Dmultiplet}
\\
\bar H^{(\bar c)}_d(\bar v) &=& \frac{1 - \bar v\!\!\!\slash}{2}
\big[ \bar D^{\mu}_d(\bar v)^\dagger \, \gamma_\mu + \bar D_d(\bar v)^\dagger\,  \gamma_5 \big] .
\label{Dbarmultiplet}
\end{eqnarray}
\label{D,Dbarmultiplet}%
\end{subequations}
The subscripts $c$ and $d$ are the isospin indices of the isospin-doublet fields.
The multiplet fields satisfy
$\bar H^{(c)}_c  v\!\!\!\slash = \bar H^{(c)}_c$ and 
$\bar v\!\!\!\slash \bar H^{(\bar c)}_d= - \bar H^{(\bar c)}_d$.
An interaction term that produces the decay $B \to K D^{(*)} \bar D^{(*)}$ must  have a factor
of the $B$-meson field $B_a$. 
It is convenient to express the field that annihilates the $B$ meson  with 4-velocity $v_B$
as a $4\times 4$ matrix obtained by setting the $B^*$ field to zero in the antibottom-meson multiplet field:
\begin{equation}
H^{(\bar b)}_a(v_B) =  \big[ - B_a(v_B)\,  \gamma_5 \big]  \frac{1 - v\!\!\!\slash_B}{2} .
\label{Bmultiplet}
\end{equation}
This field satisfies $H^{(\bar b)}_a v\!\!\!\slash_B = -H^{(\bar b)}_a$.  An interaction term 
that produces the decay $B \to K D^{(*)} \bar D^{(*)}$ must also have a factor of the kaon field $K_b^\dagger$.
The Goldstone nature of the $K$ requires the matrix element to have a factor of its 4-momentum $k^\mu$.
The Lorentz index of $k^\mu$ can be contracted with that of a Dirac matrix $\gamma^\mu$.
Lorentz-invariant interaction terms  can be expressed as Dirac traces of products of 
$H^{(\bar b)}_a(v_B)$, $\bar H^{(c)}_c(v)$, $\bar H^{(\bar c)}_d(\bar v)$
and Dirac matrices in which all Lorentz indices are contracted.

Voloshin has pointed out that the even charge conjugation of the $X(3872)$ together with the S-wave nature 
of  the dominant $D^{*0} \bar D^0$ and $D^0 \bar D^{*0}$ components of its wavefunction 
imply that the $c \bar c$ pair must be in a spin-triplet state \cite{Voloshin:2004mh}.
Since the $B$ decays into $KX$, the amplitudes for $B$ to decay into $KD^* \bar D$ and $KD \bar D^*$ must
have a substantial component in which the $c \bar c$ pair is in a spin-triplet state
near the point on the edge of the Dalitz plot where the charm mesons have equal 4-velocities. 
The simplest way to deduce the behavior of an interaction term under rotations of the heavy-quark spins
is through a nonrelativistic reduction using the methods of Ref.~\cite{Hu:2005gf}.
Interaction terms for which the $c \bar c$ pair is in a spin-triplet state can be constructed by requiring 
the charm-meson multiplet fields to appear in the combination $\bar H^{(c)}_c\gamma^\mu \bar H^{(\bar c)}_d$.
The simplest such interaction terms that are nonzero when the charm mesons have equal 4-velocities are
\begin{equation}
\frac{1}{M_B}\mathrm{Tr} \Big[ \bar H^{(c)}_c(v)\gamma^\mu \bar H^{(\bar c)}_d(v)\, 
 \Big( B_{abcd} \, H^{(\bar b)}_a(v_B)  + C_{abcd}\,\big[ H^{(\bar b)}_a(v_B),  \gamma_5 \big] \Big) \Big] 
 k_\mu K_b(k)^\dagger,
\label{SSinteraction}
\end{equation}
where the complex  coefficients $B_{abcd}$ and $C_{abcd}$ are dimensionless.
Interaction terms for which the $c \bar c$ pair is in a spin-singlet state 
when the charm mesons have equal 4-velocities can be constructed by requiring the 
charm-meson multiplet fields to appear in the combination $\bar H^{(c)}_c\gamma^5 \bar H^{(\bar c)}_d$.

Conservation of electric charge implies that there are 6 sets of subscripts 
for which the coefficients $B_{abcd}$ and $C_{abcd}$ are nonzero. 
Isospin symmetry can be used to reduce each set of coefficients $B_{abcd}$ and $C_{abcd}$ 
in Eq.~\eqref{SSinteraction} to two complex isospin coefficients
that correspond to $D^{(*)}K$ with total isospin quantum number 0 or 1.
The nonzero coefficients $B_{abcd}$ are  linear combinations of isospin coefficients $B_0$ and $B_1$ 
analogous to the linear combinations of isospin amplitudes on the right sides of Eqs.~\eqref{ABKD0D0} and \eqref{ABKD*D*},
and similarly for $C_{abcd}$.  With isospin symmetry, the  interaction terms in Eq.~\eqref{SSinteraction}
are determined by the  4  isospin coefficients $B_0$, $B_1$, $C_0$, and $C_1$. 

It is possible that the interaction terms in Eq.~\eqref{SSinteraction} for which the $c \bar c$ pair 
is in a spin-triplet state when the charm mesons have equal 4-velocities actually dominate. 
We will refer to this possibility as {\it spin-triplet dominance}.
In the  isospin analysis  in Ref.~\cite{Poireau:2011gv}, Poireau and Zito 
determined 2 constant complex amplitudes that determine 6 decay rates
for each of the 4 sets of channels $K D \bar D$, $K D^* \bar D$, $K D\bar D^*$, and $K D^* \bar D^*$.
Since one can choose phases so that one of each pair of amplitudes is real, there are 12 real parameters.
The assumption of spin-triplet dominance gives interaction terms with 4
complex isospin coefficients that determine the  amplitudes
for all the channels $K D \bar D$, $K D^* \bar D$, $K D\bar D^*$, and $K D^* \bar D^*$.
Since one coefficient can be chosen to be real, there are 7 real coefficients.
They might provide enough freedom to reproduce the  12 real parameters in  the isospin analysis
of Ref.~\cite{Poireau:2011gv} to within the errors.
The results of that isospin analysis could certainly be reproduced by adding interaction terms 
for which the $c \bar c$ pair is in a spin-singlet state when the charm mesons have equal 4-velocities.

The matrix elements that correspond to the spin-triplet interaction terms in Eq.~\eqref{SSinteraction}
can be determined by evaluating the Dirac traces.  The matrix elements are Lorentz-invariant functions
of the 4-momenta $P$, $p$, $\bar p$, $k$ (with $P=p+\bar p+k$) and the polarization 4-vectors
$\varepsilon$ and $\bar \varepsilon$ of $D^*$ and $\bar D^*$ 
(which satisfy $p \cdot \varepsilon=0$ and $\bar p \cdot \bar \varepsilon=0$).
At the point on the edge of the Dalitz plot where the charm mesons have equal 4-velocities,
the matrix elements reduce to
\begin{subequations}
\begin{eqnarray}
\mathcal{A}[B \to K D \bar D]&=&
-C\frac{\lambda(M_B,2M_D,m_K)}{8 M_B^2 M_D^2},
\label{ASS-DDbar}
\\
\mathcal{A}[B \to K D^* \bar D] &=& 
- B\frac{(M_B+M_{D^*}+M_D)^2 - m_K^2}{2 M_B^2 (M_{D^*}+M_D)} \, P \!\cdot\! \varepsilon ,
\label{ASS-D*Dbar}
\\
\mathcal{A}[B \to K D \bar D^*]&=&
- B\frac{(M_B+M_{D^*}+M_D)^2 - m_K^2}{2 M_B^2 (M_{D^*}+M_D)} \, P \!\cdot\! \bar\varepsilon ,
\label{ASS-DDbar*}
\\
\mathcal{A}[B \to K D^* \bar D^*]&=&
-  C \,\left( \frac{\lambda(M_B,2M_{D^*},m_K)}{8 M_B^2 M_{D^*}^2}\,  \varepsilon\! \cdot \! \bar\varepsilon 
+ \frac{4}{M_B^2} \, P \!\cdot\! \varepsilon P \!\cdot \!\bar \varepsilon \right)
\nonumber\\
&& 
+i B \,  \frac{(M_B+2M_{D^*})^2 - m_K^2}{8M_B^2 M_{D^*}^2} \, 
\epsilon_{\mu\nu\alpha\beta}P^\mu k^\nu \varepsilon^\alpha \bar \varepsilon^\beta.
\label{ASS-D*Dbar*}
\end{eqnarray}
\label{ASS}%
\end{subequations}
On the left side, we have suppressed the isospin indices $a$, $b$, $c$, and $d$ of the mesons
$B$, $K$, $D^{(*)}$, and $\bar D^{(*)}$.
On the right side, we have suppressed the subscripts $abcd$ of the coefficients $B$ and $C$.
The nonzero coefficients can be expressed in terms of isospin coefficients $B_0$, $B_1$, $C_0$, and $C_1$.


\section{Decays into  $\bm{K}$ Plus $\bm{D^* \bar{D}^*}$ near Threshold}
\label{sec:BtoKD*D*0}

\begin{figure}[t]
\includegraphics*[width=0.3\linewidth]{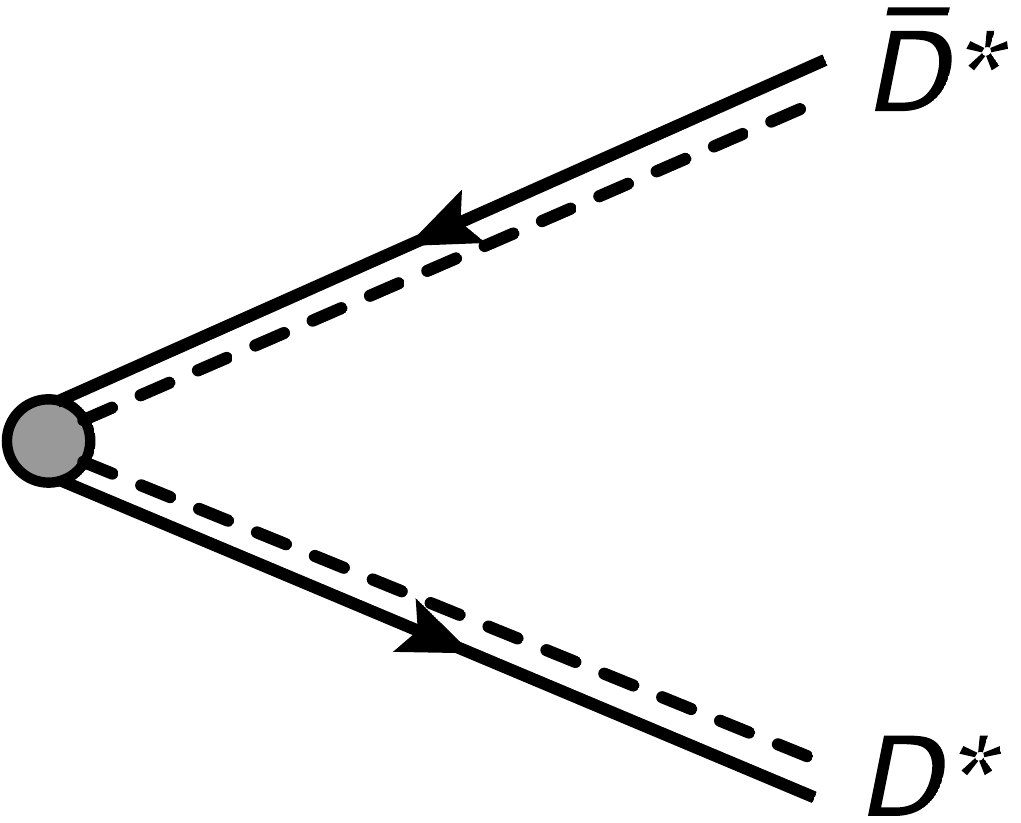} 
\caption{
Feynman diagram in XEFT for production of $D^* \bar D^*$ from their creation at a point.
The $D^*$ and $\bar D^*$ are represented by double lines consisting of  a dashed line
and a solid  line with an arrow. 
}
\label{fig:D*D*}
\end{figure}

In the decay of a $B$ meson into $KD^* \bar D^*$ with the pair of spin-1 charm mesons 
having small relative momentum, the momentum in the charm-meson-pair rest frame 
of either the incoming $B$ or the  outgoing $K$  is  about 1350~MeV. 
Since this is much larger than $m_\pi$,
the $B$-to-$K$ transition occurs over distances much shorter than the range $1/m_\pi$
of the interactions between the charm mesons.
As far as the $D^*$ and $\bar D^*$ are concerned, the $B \to K$ transition can be described 
as a point interaction that creates $D^*$ and $\bar D^*$. 
The amplitude for producing $D^* \bar D^*$ can be represented in XEFT by  
the Feynman diagram in Fig.~\ref{fig:D*D*} with a vertex from which the $D^*$ and $\bar D^*$ emerge.
The vertex factor for the $B$-to-$K$ transition that creates $D^* \bar D^*$ at a point is
$i \mathcal{A}^{ij}[KD^* \bar D^*]$, where $i$ and $j$ are the spin indices of the $D^*$ and $ \bar D^*$.
The vertex factor $\mathcal{A}^{ij}$ is a Cartesian tensor in the center-of-momentum (CM) frame 
of $D^* \bar D^*$.  The only preferred direction is that of the 3-momentum $\bm{P}$
of the decaying $B$ meson, which is also the direction of the 3-momentum of the final-state $K$.
The amplitude must therefore have the tensor structure
\begin{equation}
\mathcal{A}^{ij}\big[B \to KD^* \bar D^*\big] =
D\,  \delta^{ij}   + E\, \hat P^i  \hat P^j + i F  \,\epsilon^{ijk}\hat P^k,
\label{Aij}
\end{equation}
where the  complex coefficients $D$, $E$, and $F$ are dimensionless.
We have suppressed the isospin indices $a$, $b$, $c$, and $d$ of the mesons
$B$, $K$, $D^{(*)}$, and $\bar D^{(*)}$ and the subscripts $abcd$ of the coefficients  $D$, $E$, and $F$.
The nonzero coefficients   $D_{abcd}$ can be expressed as linear combinations 
of two isospin coefficients  $D_0$ and $D_1$ analogous to 
the linear combinations in Eqs.~\eqref{ABKD*D*}, and similarly for $E_{abcd}$ and $F_{abcd}$.
If we  make the approximation of spin-triplet dominance 
that gives the Lorentz-invariant amplitude in Eq.~\eqref{ASS-D*Dbar*}, the coefficients are
\begin{subequations}
\begin{eqnarray}
D_i &\approx& C_i \,\frac{\lambda(M_B,2M_{D^*},m_K)}{8 M_B^2 M_{D^*}^2},
\label{Di}
\\
E_i &\approx& -C_i \,\frac{\lambda(M_B,2M_{D^*},m_K)}{4 M_B^2 M_{D^*}^2},
\label{Ei}
\\
F_i&\approx& B_i \,  \frac{\big[ (M_B+2M_{D^*})^2 - m_K^2 \big]\, \lambda^{1/2}(M_B,2M_{D^*},m_K)}{16 M_B^2 M_{D^*}^2} .
\label{Fi}
\end{eqnarray}
\label{abctrip}%
\end{subequations}
Note that the assumption of spin-triplet dominance implies $E_i =-2D_i$.

The matrix element for producing  $D^* \bar D^*$ is obtained by contracting the tensor $\mathcal{A}^{ij}[KD^* \bar D^*]$
with the polarization vectors $\varepsilon^i$ and $\bar \varepsilon^j$ of the $D^*$ and $\bar D^*$.
If the amplitude $\mathcal{A}^{ij}$ in Eq.~\eqref{Aij} is contracted with $\varepsilon^i\bar \varepsilon^j$,
multiplied by its complex conjugate,
and then summed over the spin states of $D^*$ and $\bar D^*$, the result  is
\begin{equation}
\big|\mathcal{A}[K D^* \bar D^*] \big|^2 \equiv
\sum_\mathrm{spins} \big| \varepsilon^i  \, \mathcal{A}^{ij} \, \bar \varepsilon^j \big|^2 =
2 \, | D |^2 +  | D +E |^2 + 2|F |^2.
\label{sumAA*}
\end{equation}
We use the symbol $|\mathcal{A}[K D^* \bar D^*]|^2$ as a concise notation for the sum in Eq.~\eqref{sumAA*},
and we refer to it as a {\it squared amplitude},  even though it is actually a sum of squares.
For any specific decay channel $KD^* \bar D^*$, $D$ can be expressed as the same linear combination 
of isospin coefficients $D_0$ and $D_1$ as in Eqs.~\eqref{ABKD*D*}, and similarly for  $E$ and $F$.
For the decays  that produce $D^{*0}$, the resulting expressions for the  squared amplitudes  are
\begin{subequations}
\begin{eqnarray}
\left|\mathcal{A}\big[K^0 D^{*0} \bar D^{*0}\big]  \right|^2&=& 
\frac{2\big( 2 \big|D_1 \big|^2 + \big|D_1 + E_1 \big|^2  + 2 \big|F_1 \big|^2\big)}{3},
\label{AsqK0D*0D*0}
\\
\left|\mathcal{A}\big[K^+  D^{*0} D^{*-}\big]  \right|^2&=& 
\frac{2 \big|D_1 -\sqrt{3}  D_0 \big|^2 + \big|D_1 + E_1 -\sqrt{3}  (D_0  +  E_0) \big|^2  
+ 2 \big|F_1 - \sqrt{3}  F_0\big|^2}{6},~~~~
\label{AsqK+D*0D*-}
\\
\left|\mathcal{A}\big[K^+  D^{*0} \bar D^{*0}\big]  \right|^2&=& 
\frac{2 \big|D_1 + \sqrt{3}  D_0 \big|^2 + \big|D_1 + E_1 + \sqrt{3}  (D_0  +  E_0) \big|^2  
+ 2 \big|F_1 + \sqrt{3}  F_0\big|^2}{6}.~~~~
\label{AsqK+D*0D*0}
\end{eqnarray}
\label{AsqKD*D*0}%
\end{subequations}
These squared amplitudes depend on 10 independent real components of the 6 isospin coefficients.
The squared amplitude  $|\mathcal{A}[K^0  D^{*+} \bar D^{*0}]|^2$
is equal to $|\mathcal{A}[K^+  D^{*0} D^{*-} ] |^2$ by isospin symmetry.

The differential decay rates for producing $D^* \bar D^*$ with small relative momentum can be obtained
by multiplying the squared amplitudes, such as those in Eqs.~\eqref{AsqKD*D*0},
by the differential phase space for $K D^* \bar D^*$ and by $1/2M_B$.
If  the amplitudes do not vary dramatically across the Dalitz plot, the resulting expressions 
for the differential decay rates may also be reasonable approximations throughout the Dalitz plot.
In this case, the squared amplitudes in Eqs.~\eqref{AsqKD*D*0} can be approximated by the 
squares of the corresponding amplitudes in Eqs.~\eqref{ABKD*D*}.
The isospin amplitudes $A_0^{**}$ and $A_1^{**}$ from the isospin analysis of 
 Poireau and Zito in Ref.~\cite{Poireau:2011gv} are given by the last row of Table~\ref{tab:PZamps}.
 Inserting them into the amplitudes in Eqs.~\eqref{AB0K0D*0D*0}, \eqref{AB0K+D*0D*-}, and \eqref{ABpK+D*0D*0},
and then evaluating their absolute squares, we obtain
\begin{subequations}
\begin{eqnarray}
\Big| \mathcal{A}\big[K^0 D^{*0} \bar D^{*0}\big]\Big|^2 &=& 
(0.35 \pm 0.05) \times 10^{-10} ,
\label{AsqK0D*0D*0num}
\\
\Big| \mathcal{A}\big[K^+  D^{*0} D^{*-}\big] \Big|^2 &=& 
(2.85 \pm 0.22) \times 10^{-10} ,
\label{AsqK+D*0D*num}
\\
\Big| \mathcal{A}\big[K^+  D^{*0} \bar D^{*0}\big] \Big|^2 &=& 
(2.52 \pm 0.21) \times 10^{-10} .
\label{AsqK+D*0D*0num}
\end{eqnarray}
\label{AsqKD*D*num}%
\end{subequations}
These equations provide 3 real constraints on the 
6 complex isospin coefficients in Eqs.~\eqref{AsqKD*D*0}.

The squared amplitudes in Eqs.~\eqref{AsqKD*D*0} can be simplified by assuming
the spin-triplet dominance of the amplitudes, which implies $E_i=-2D_i$.
They then reduce to 
\begin{subequations}
\begin{eqnarray}
\left|\mathcal{A}\big[K^0 D^{*0} \bar D^{*0}\big]  \right|^2&\approx& 
2 \big|D_1 \big|^2  + \tfrac43 \big|F_1 \big|^2,
\label{AsqK0D*0D*0-STD}
\\
\left|\mathcal{A}\big[K^+  D^{*0} D^{*-}\big]  \right|^2&\approx& 
\tfrac12   \big|D_1 -\sqrt{3}  D_0 \big|^2 + \tfrac13 \big|F_1 - \sqrt{3}  F_0\big|^2,
\label{AsqK+D*0D*-STD}
\\
\left|\mathcal{A}\big[K^+  D^{*0} \bar D^{*0}\big]  \right|^2&\approx& 
\tfrac12   \big|D_1  +\sqrt{3}  D_0 \big|^2 +\tfrac13 \big|F_1 + \sqrt{3}  F_0\big|^2.
\label{AsqK+D*0D*0-STD}
\end{eqnarray}
\label{AsqKD*D*-STD}%
\end{subequations}
These squared amplitudes depend on 6 independent real  components of the 6 complex isospin coefficients.
The values of the squared amplitudes in Eqs.~\eqref{AsqKD*D*num} provide 3 real constraints.


\section{Decays into  $\bm{K}$ Plus $\bm{X}$ and a  Pion}
\label{sec:BtoKXpi}

\begin{figure}[t]
\includegraphics*[width=0.4\linewidth]{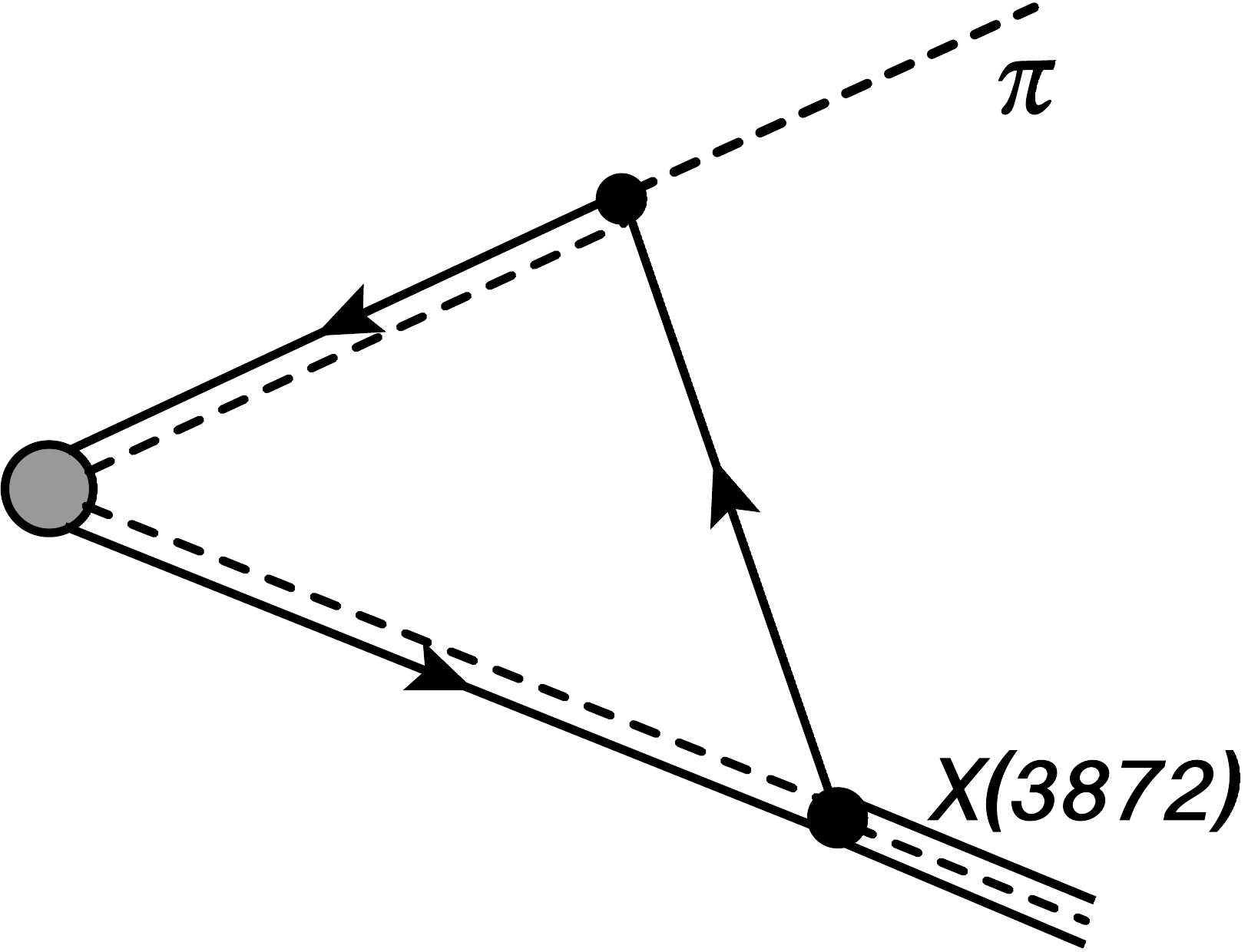} ~
\includegraphics*[width=0.4\linewidth]{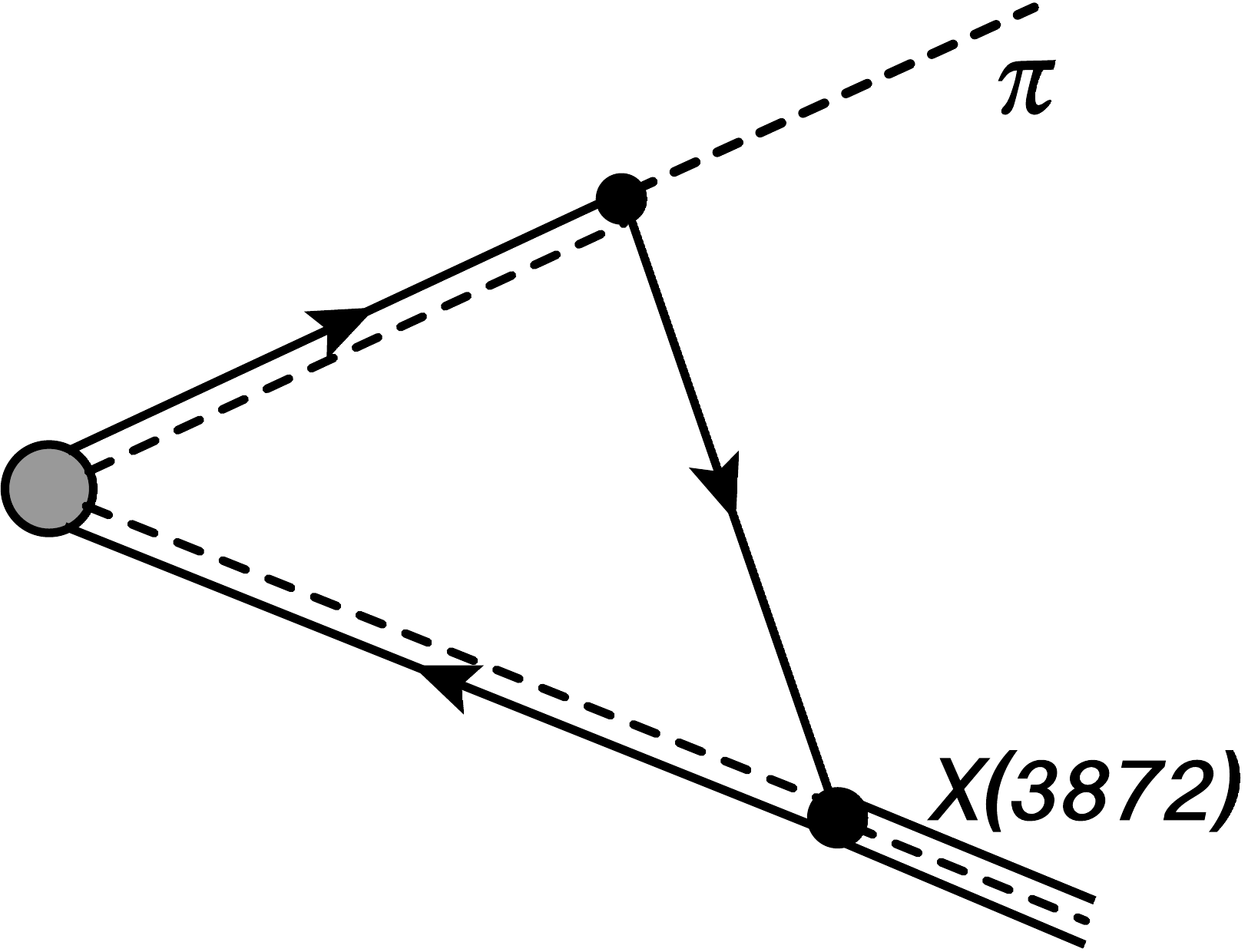} 
\caption{
Feynman diagrams in XEFT for $D^* \bar D^*$ created at a point to rescatter into $X\pi$.
The $D$ and $\bar D$ are represented by solid lines with an arrow.
The $X$ is represented by a triple line consisting of two solid lines and one dashed line.
The $\pi$ is represented by a dashed line.
}
\label{fig:DDtoXpi}
\end{figure}

A pair of spin-1 charm mesons $D^* \bar D^*$ created at short distances with relative momentum $\bm{k}$
can rescatter into $X(3872)\pi$  with relative momentum $\bm{q}$. 
The rescattering  can be described within XEFT provided the relative momentum  
of the charm mesons that form the $X$ is less than about $m_\pi$. 
The Feynman diagrams for $D^* \bar D^*$ created at a point to rescatter 
into $X\pi$ are shown in Fig.~\ref{fig:DDtoXpi}.
These diagrams can be calculated using the Feynman rules for Galilean-invariant XEFT in Ref.~\cite{Braaten:2015tga} 
together with the vertices for the coupling of $D^{*0} \bar D^0$ and $D^{0} \bar D^{*0}$ to $X$ in Ref.~\cite{Braaten:2010mg}.
These vertices are given by $(\sqrt{\pi \gamma_X}/\mu) \delta^{ij}$, 
where $i$ and $j$ are the spin indices of the  spin-1 charm meson and the $X$.

In the case of the production of $X \pi^0$  from $D^{*0} \bar D^{*0}$ created at short distances,
the amplitude is given by the sum of the two diagrams in   Fig.~\ref{fig:DDtoXpi}.
The integral over the loop energy is conveniently evaluated by contours using the pole of the 
propagator for the $D^*$ or $\bar D^*$ line attached to the $X$.
The remaining two propagators can be combined into a single denominator by introducing 
an integral over a Feynman parameter $x$.  
The integral over the loop momentum can  be evaluated analytically.
Our result for the amplitude for producing $X \pi^0$ with small relative momentum $\bm{q}$
and with polarization vector $\bm{\varepsilon}$ for the $X$ is
\begin{eqnarray}
&&i\, \mathcal{A}^{ij}[KD^{*0} \bar D^{*0}]
 \frac{g (\pi \gamma_X M_{*0}^3 /m_0)^{1/2}}{16 \pi  \mu f_\pi}
 \big(\varepsilon^i q^j + q^i \varepsilon^j \big)
 \int_0^1 dx \left( \frac{2M_0}{2M_0+(1-x)m_0} \right)^{5/2}
\nonumber\\
 && \hspace{1.5cm} \times
 \Bigg[ \big(\delta_0-\gamma_X^2/2\mu\big) - (1+x)\big(\delta_0-i \Gamma_{*0}/2\big)
+\frac{M_0x}{(2M_0+(1-x)m_0) \mu_{X \pi} }  \bm{q}^2 \Bigg]^{-1/2},~~~~~~
\label{amplitudeXpi-1}
\end{eqnarray}
where $\delta_0=M_{*0} - M_0- m_0 = 7.0$~MeV, 
$\Gamma_{*0} \approx 60$~keV is the predicted decay width of $D^{*0}$ \cite{Braaten:2015tga},
and $\mu_{X \pi} = ((2M_0+m_0)m_0)/(2(M_0+m_0))$ is the Galilean-invariant reduced mass of $X\pi$.
The coupling constant for the pion-emission vertex is $g/(2\sqrt{m_0} f_\pi)= 0.30/m_0^{3/2}$ \cite{Braaten:2015tga}.
The final integral over $x$ can also be evaluated analytically
if the integrand is simplified using $m_0 \ll M_0$. 
Our final result for the amplitude  is relatively simple:
\begin{equation}
i\mathcal{A}^{ij}[KD^{*0} \bar D^{*0}]
 \frac{g (\pi \gamma_X M_{*0}^3/m_0)^{1/2}}{8 \pi  \mu f_\pi} 
  \frac{\varepsilon^i q^j + q^i \varepsilon^j }{\sqrt{q^2/2m_0  - \delta_0 -\gamma_X^2/2\mu + i \Gamma_{*0}} + \sqrt{-\gamma_X^2/2\mu + i \Gamma_{*0}/2}}.
\label{amplitudeXpi0}
\end{equation}
The denominator is a kinematic singularity factor that would have a zero at  $q^2= 2 m_0 \delta_0$
if the binding momentum $\gamma_X$ and the width $\Gamma_{*0}$ were both zero.
The kinematic singularity is called a {\it triangle singularity},
because it arises from the region where the three charm meson lines that form a triangle 
in the Feynman diagrams in Fig.~\ref{fig:DDtoXpi} are all simultaneously on shell.

In the case of the production of  $X \pi^-$  from $D^{*0} D^{*-}$ created at short distances,
the amplitude is given by the  first diagram in Fig.~\ref{fig:DDtoXpi} only.  
The coupling constant for the pion-emission vertex is  $g/(\sqrt{2m_0} f_\pi)$. 
If the loop integral is simplified using $m_0 \ll M_0$, our final result for the amplitude
for producing $X \pi^-$ with small relative momentum $\bm{q}$
and with polarization vector $\bm{\varepsilon}$ for the $X$  is
\begin{eqnarray}
&&i \mathcal{A}^{ij}[KD^{*0} D^{*-}]
 \frac{g (2\pi \gamma_X M_{*0}^3/m_0)^{1/2}}{8 \pi  \mu f_\pi} 
\nonumber\\
&& \hspace{2cm} \times
\frac{\varepsilon^i q^j}{\sqrt{q^2/2m_0  - \delta_1 -\gamma_X^2/2\mu + i (\Gamma_{*0} +  \Gamma_{*1})/2} 
  + \sqrt{-\gamma_X^2/2\mu + i \Gamma_{*0}/2}},
\label{amplitudeXpi+}
\end{eqnarray}
where $\delta_1=M_{*1} - M_0-m_1= 5.9$~MeV
and $\Gamma_{*1} \approx 83$~keV is the measured decay width of $D^{*-}$.
The amplitude for producing $X \pi^+$  from $D^{*+} \bar D^{*0}$ created at short distances 
can be obtained by replacing the vertex factor  by $\mathcal{A}^{ij}[KD^{*+} \bar D^{*0}]$
and replacing $\varepsilon^i q^j$ by $q^i \varepsilon^j$.
The denominator of Eq.~\eqref{amplitudeXpi+} is a triangle-singularity factor 
that would have a zero at  $q^2= 2 m_0 \delta_1$
if the binding momentum $\gamma_X$ and the widths $\Gamma_{*0}$ and $\Gamma_{*1}$ were all zero.

Our expressions for the amplitudes  in Eqs.~\eqref{amplitudeXpi0} and \eqref{amplitudeXpi+}
should be accurate provided the  momentum integral that results in Eq.~\eqref{amplitudeXpi-1} 
is dominated by regions where the relative momentum $\bm{k}$ of the charm mesons that form the $X$ 
is less than about $m_\pi$. 
This condition imposes a constraint on the relative momentum $\bm q$ of $X$ and $\pi$.
The  constraint can be deduced from the integrated expression in Eq.~\eqref{amplitudeXpi-1}  
by requiring the energy proportional to $\bm q^2$ inside the last factor to be less than $m_\pi^2/2\mu$.
At $x=1$, this energy is $E_{X\pi} = q^2/2\mu_{X\pi}$. 
Thus the kinetic energy $E_{X\pi}$ must be less than about $m_\pi^2/2\mu \approx 10$~MeV.

To obtain the differential decay rate for producing $X\pi$ with small relative momentum,
the amplitude in Eq.~\eqref{amplitudeXpi0} or \eqref{amplitudeXpi+} must be multiplied  by its complex conjugate,
summed over the spin states of $X$,
and then multiplied by the differential phase space for $K X \pi$ and by $1/2M_B$.
The  differential decay rate for producing $X\pi$ with relative momentum $\bm{q}$ is
\begin{equation}
d\Gamma[B \to K X\pi] =
\frac{1}{2 M_B}  \int d\Phi_{(D^*\bar D^*)K}
\Big| \mathcal{A}[K X\pi]\Big|^2 \frac{d^3q}{(2 \pi)^32 \mu_{X\pi}},
\label{dGammaBtoKXpi}
\end{equation}
where $d\Phi_{(D^*\bar D^*)K}$ is the  differential two-body phase space 
for $K$ and a composite particle denoted by  $(D^*\bar D^*)$ with mass $2M_{D^*}$.
The  differential decay rate can be simplified by averaging over the directions of $\bm{q}$
or, equivalently, by averaging over the directions of the momentum $\bm{P}$ of $B$.
The average of the product of the amplitude $\mathcal{A}^{ij}$ in Eq.~\eqref{Aij}
and its complex conjugate $(\mathcal{A}^{kl})^*$ over the directions of the momentum of the $B$ is
\begin{equation}
\big\langle \mathcal{A}^{ij} \big(\mathcal{A}^{kl}\big)^* \big\rangle
=  \big| D + \tfrac13 E \big|^2 \delta^{ij} \delta^{kl}  
+ \tfrac{1}{15} |E|^2\, \big( \delta^{ik} \delta^{jl}  +\delta^{il} \delta^{jk} - \tfrac23  \delta^{ij} \delta^{kl}  \big)
+ \tfrac{1}{3} |F|^2 \, \big(\delta^{ik} \delta^{jl} - \delta^{il} \delta^{jk} \big).
\label{AASS-D*Dbar*}
\end{equation}
If the amplitude $\mathcal{A}^{ij}$ in Eq.~\eqref{Aij} is contracted with the tensors in the numerators of Eqs.~\eqref{amplitudeXpi0} and \eqref{amplitudeXpi+},   multiplied by its complex conjugate, 
and then summed over the spin states of $X$, the results are
\begin{subequations}
\begin{eqnarray}
&&\sum_\mathrm{spins} \big\langle\mathcal{A}^{ij} \big(\mathcal{A}^{kl}\big)^* \big\rangle
(\varepsilon^i q^j + q^i \varepsilon^j ) (\varepsilon^k q^l + q^k \varepsilon^l )^*
=   4 \left( \big| D + \tfrac13 E \big|^2 + \tfrac{2}{9} |E|^2 \right) \bm{q}^2,
\label{sumAASS-D*Dbar*0}
\\
&&\sum_\mathrm{spins}\big\langle\mathcal{A}^{ij} \big(\mathcal{A}^{kl}\big)^*\big\rangle
(\varepsilon^i q^j)  (\varepsilon^k q^l)^*
=   \left( \big| D + \tfrac13 E \big|^2 + \tfrac{2}{9} |E|^2 + \tfrac{2}{3} |F|^2 \right) \bm{q}^2.
\label{sumAASS-D*Dbar*pm}
\end{eqnarray}
\label{sumAASS}%
\end{subequations}
For any specific decay channel $KX\pi$, the coefficient $D$ can be expressed as the same linear combination 
of isospin coefficients $D_0$ and $D_1$ as in Eqs.~\eqref{ABKD*D*}, and similarly for  $E$ and $F$.
We introduce a compact notation for the factors that depend on the isospin coefficients:
\begin{subequations}
\begin{eqnarray}
\left|\mathcal{A}\big[K^0 X \pi^0 \big]  \right|^2&\equiv& 
\big|D_1 + \tfrac13 E_1 \big|^2  + \tfrac29 \big|E_1 \big|^2,
\label{AsqK0Xpi0}
\\
\left|\mathcal{A}\big[K^+  X \pi^- \big]  \right|^2 &\equiv& 
 \big|D_1 + \tfrac13 E_1-\sqrt{3}  (D_0  +  \tfrac13 E_0) \big|^2  
+ \tfrac29 \big|E_1 - \sqrt{3}  E_0\big|^2 + \tfrac23 \big|F_1 - \sqrt{3}  F_0\big|^2,
\label{AsqK+Xpi-}
\\
\left|\mathcal{A}\big[K^+  X \pi^0 \big]  \right|^2&\equiv& 
\big|D_1 + \tfrac13 E_1+\sqrt{3}  (D_0  +  \tfrac13E_0) \big|^2  + \tfrac29 \big|E_1 + \sqrt{3}  E_0\big|^2,
\label{AsqK+Xpi0}
\\
\left|\mathcal{A}\big[K^0  X \pi^+ \big]  \right|^2&\equiv& 
\big|D_1 + \tfrac13 E_1-\sqrt{3}  (D_0  +  \tfrac13 E_0) \big|^2  
+ \tfrac29 \big|E_1 - \sqrt{3}  E_0\big|^2 + \tfrac23 \big|F_1 - \sqrt{3}  F_0\big|^2.~~~~~
\label{AsqK0Xpi+}
\end{eqnarray}
\label{AsqKXpi}%
\end{subequations}
Note that $|\mathcal{A}[K^0  X \pi^+ ]|^2$ is equal to $|\mathcal{A}[K^+  X \pi^- ] |^2$.
We refer to these factors as squared amplitudes, even though they are actually sums of squares.

The differential  decay rates for $B^0$ into $K^0  X \pi^0$ and into $K^+  X \pi^-$
with small relative momentum $\bm{q}$ for  $X\pi$ are
\begin{subequations}
\begin{eqnarray}
\frac{d\Gamma}{d^3q}[B^0 \to  K^0  X \pi^0]
&=& 
\left|\mathcal{A}\big[K^0 X \pi^0 \big]  \right|^2 \frac{g^2 \,  \lambda^{1/2}(M_B , 2M_{D^*},m_K ) \,M_{D^*}^3\gamma_X}
{96(2 \pi)^5  M_B^3\, \mu^2  f_\pi^2 \,\mu_{X\pi}} 
 \nonumber\\ 
&&  \hspace{-2cm} \times
\frac{q^2/2m_0}
{\big| \sqrt{q^2/2m_0  - \delta_0 -\gamma_X^2/2\mu + i \Gamma_{*0}} + \sqrt{-\gamma_X^2/2\mu + i \Gamma_{*0}/2}  \big|^2},
\label{dGammaB+toKpXpi0}
\\
\frac{d\Gamma}{d^3q}[B^0 \to  K^+  X \pi^-]
&=&
\left|\mathcal{A}\big[K^+  X \pi^- \big]  \right|^2
\frac{g^2 \,  \lambda^{1/2}(M_B ,  2M_{D^*},m_K )\, M_{D^*}^3\gamma_X}
{768(2 \pi)^5  M_B^3\, \mu^2  f_\pi^2\,  \mu_{X\pi}} 
 \nonumber\\ 
&& \hspace{-2cm} \times
\frac{q^2/2m_0}
{\big| \sqrt{q^2/2m_0  - \delta_1  -\gamma_X^2/2\mu+ i (\Gamma_{*0}+ \Gamma_{*1})/2}  
             +  \sqrt{ -\gamma_X^2/2\mu + i \Gamma_{*0}/2} \big|^2}.
\label{dGammaB+toK0Xpi+}
\end{eqnarray}
\label{dGammaB+toKXpi}%
\end{subequations}
The differential decay rate for $B^+ \to  K^+  X \pi^0$ differs from that for $B^0 \to  K^0  X \pi^0$ only
by an overall multiplicative factor that depends on isospin coefficients,
while the differential decay rate for 
$B^+ \to  K^0  X \pi^+$ is the same as that for $B^0 \to  K^+  X \pi^-$:
\begin{subequations}
\begin{eqnarray}
\frac{d\Gamma}{d^3q}[B^+ \to  K^+  X \pi^0]
&=&  
\frac{\left|\mathcal{A}\big[K^+  X \pi^0 \big]  \right|^2}{4\left|\mathcal{A}\big[K^0  X \pi^0 \big]  \right|^2}~
\frac{d\Gamma}{d^3q}[B^0 \to  K^0  X \pi^0],
\label{dBrB0toXpi0}
\\
\frac{d\Gamma}{d^3q}[B^+ \to  K^0  X \pi^+]
&=& \frac{d\Gamma}{d^3q}[B^0 \to  K^+  X \pi^-].
\label{dBrB0toXpi-}
\end{eqnarray}
\label{dBrB0toXpi}%
\end{subequations}

\begin{figure}[t]
\includegraphics*[width=0.8\linewidth]{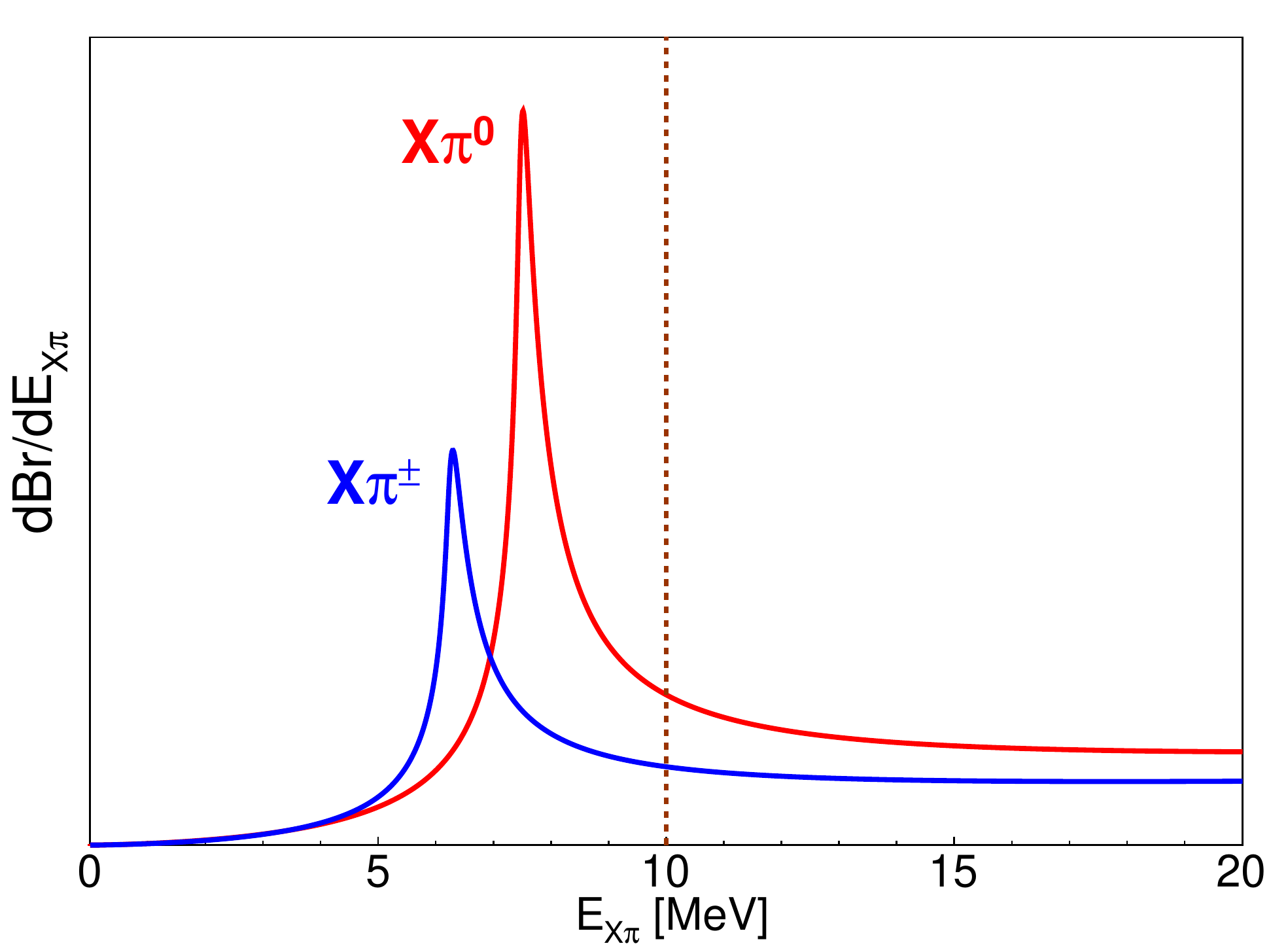} 
\caption{
Differential branching fractions $d\mathrm{Br}/dE_{X\pi}$
for the decays $B^0 \to K^0 X \pi^0$ and $B^+ \to K^+ X \pi^0$
 (taller red curve) and $B^0 \to K^+ X \pi^-$ and  $B^+ \to K^0 X \pi^+$ (shorter blue curve)
as functions of the kinetic energy $E_{X\pi} = q^2/2\mu_{X\pi}$ of $X\pi$ in its CM frame.
The binding energy of the $X$ is 0.17~MeV.
The region of validity of XEFT extends out to about the vertical dotted line at $E_{X\pi} = 10$~MeV. 
The  normalizations of the curves are arbitrary.
The relative normalizations of the $X \pi^0$ and $X\pi^\pm$ curves are chosen so their extrapolations to
 large $E_{X\pi}$ are equal.}
\label{fig:Xpi-q}
\end{figure}

The differential  decay rates in Eqs.~\eqref{dGammaB+toKXpi}
can be expressed as differential branching fractions $d\mathrm{Br}/dE_{X\pi}$ 
in the kinetic energy $E_{X\pi}= q^2/2\mu_{X\pi}$ of $X$ and $\pi$ in their CM frame.
Their normalizations depend on the undetermined coefficients
$D_i$, $E_i$, and $F_i$, but their dependence on $E_{X\pi}$ is predicted.
The shapes of the differential branching fractions for the decays of $B^0$ into $K^0 X \pi^0$ and $ K^+ X \pi^-$
are illustrated in Fig.~\ref{fig:Xpi-q} for $X$ with binding energy  0.17~MeV.
The  normalizations of the curves are arbitrary.
Both of the curves have a narrow peak from a charm-meson triangle singularity.
The $B \to K$ transition creates a pair of charm mesons $D^* \bar D^*$ that are almost on shell,
one of them decays into $D \pi$ or $\bar D \pi$, 
and the resulting pair of almost on-shell charm mesons  binds to form the $X$.
For the decay $B^0 \to K^0 X \pi^0$, there is a  narrow peak in $E_{X\pi}$ near $\delta_0= 7.0$~MeV. 
The peak is produced by  the denominator in Eq.~\eqref{dGammaB+toKpXpi0}. 
The full width at half maximum of that factor is  $1.17 \, \gamma_X^2/2\mu$
if the binding energy $\gamma_X^2/2\mu$  is large compared to $\Gamma_{*0}$
and $6.21 \, \Gamma_{*0} \approx 370$~keV  
if $\gamma_X^2/2\mu$ is small compared to $\Gamma_{*0}$.
For the decay $B^0 \to K^+ X \pi^-$, there is a  narrow  peak in $E_{X\pi}$ near $\delta_1= 5.9$~MeV.
The peak is produced by the denominator in Eq.~\eqref{dGammaB+toK0Xpi+}.
The full width at half maximum of  that factor 
in Eq.~\eqref{dGammaB+toK0Xpi+} is $1.17 \, \gamma_X^2/2\mu$
if $\gamma_X^2/2\mu$  is large compared to $\Gamma_{*0}$  and $\Gamma_{*1}$ and approximately 430~keV
if $\gamma_X^2/2\mu$  is small compared to $\Gamma_{*0}$ and $\Gamma_{*1}$.

At energies above the narrow peaks,
our expressions for the differential branching fractions $d\mathrm{Br}/dE_{X\pi}$ in
Eqs.~\eqref{dGammaB+toKpXpi0} and \eqref{dGammaB+toK0Xpi+}
have local minima at energies $E_{X\pi}$ near $3 \delta_0$ and $3  \delta_1$,  respectively.
At higher energies, the distributions increase as $E_{X\pi}^{1/2}$.
This differs from the behavior $E_{X\pi}^{3/2}$ expected from the P-wave coupling of the pion
because of the resonance factors in the denominators of the amplitudes in Eqs.~\eqref{dGammaB+toKXpi}.
The region of $E_{X\pi}$ where the distributions increase is beyond the energy  $m_\pi^2/2\mu$ where XEFT breaks down, 
which is  marked by a vertical dotted line in Fig.~\ref{fig:Xpi-q}.

The contributions of the triangle singularities to the integrated decay rates 
can be estimated by integrating the momentum distributions in Eqs.~\eqref{dGammaB+toKXpi}
from the threshold to some energy  $E_\mathrm{max} = q_\mathrm{max}^2/2\mu_{X\pi}$ beyond the peak.
In the limits $\Gamma_{*0} \to 0$ and $\gamma_X \ll \sqrt{\mu \delta_0}$,
the integral of the momentum dependent factor in Eq.~\eqref{dGammaB+toKpXpi0}
over the region   $|\bm q| <  q_\mathrm{max}$ is
\begin{eqnarray}
&&\int_{q < q_\mathrm{max}}  \frac{d^3q}{(2\pi)^3} \frac{q^2/2m_0}
{\big| \sqrt{q^2/2m_0  - \delta_0 -\gamma_X^2/2\mu + i \epsilon} + i \sqrt{\gamma_X^2/2\mu}  \big|^2}
= \frac{1}{2 \pi^2} (2m_0 \delta_0)^{3/2} 
\nonumber\\
&& \hspace{0.5cm}
\times \left[  \log \frac{8 \mu\delta_0}{\gamma_X^2} + \frac13 \left( \frac{q_\mathrm{max}^2}{2m_0\delta_0} \right)^{3/2} 
+\left( \frac{q_\mathrm{max}^2}{2m_0\delta_0} \right)^{1/2}
- \frac12 \log \frac{\sqrt{q_\mathrm{max}^2/2m_0\delta_0} +1}{\sqrt{q_\mathrm{max}^2/2m_0\delta_0}-1}  -  \frac{11}{3} \right].
\label{intsoftpi}
\end{eqnarray}
The coefficient of $(2m_0 \delta_0)^{3/2}$ diverges logarithmically as $\gamma_X \to 0$.
If we do not take the limit $\Gamma_{*0} \to 0$,
the coefficient of $(m_0 \delta_0)^{3/2} $ also depends logarithmically on $\Gamma_{*0}$.
In the limit $\Gamma_{*0} \to 0$ and $\Gamma_{*1} \to 0$,
the integral of the momentum dependent factor in Eq.~\eqref{dGammaB+toK0Xpi+} is
given by  Eq.~\eqref{intsoftpi} with $\delta_0$ replaced by $\delta_1$.

The normalizations factors in the differential  decay rates in Eqs.~\eqref{dGammaB+toKXpi}
and \eqref{dBrB0toXpi} depend on the unknown isospin coefficients $D_i$, $E_i$, and $F_i$.
The normalization factors can be simplified by assuming the spin-triplet dominance of the amplitudes, 
which implies $E_i=-2D_i$.
The  three distinct squared amplitudes in Eqs.~\eqref{AsqKXpi} then reduce to 
\begin{subequations}
\begin{eqnarray}
\left|\mathcal{A}\big[K^0 X \pi^0 \big]  \right|^2 &\approx& \big|D_1 \big|^2 ,
\label{AsqK0Xpi0:STD}
\\
\left|\mathcal{A}\big[K^+  X \pi^- \big]  \right|^2 &\approx& 
\big|D_1 -\sqrt{3}  D_0 \big|^2 + \tfrac23 \big|F_1 - \sqrt{3}  F_0\big|^2,
\label{AsqK+Xpi-:STD}
\\
\left|\mathcal{A}\big[K^+  X \pi^0 \big]  \right|^2 &\approx& \big|D_1  +\sqrt{3}  D_0 \big|^2 .
\label{AsqK+Xpi0:STD}
\end{eqnarray}
\label{AsqKXpi:STD}%
\end{subequations}
These expressions  are related in a simple way to the corresponding expressions for the squared amplitudes for
$B \to K D^* \bar D^*$ in Eqs.~\eqref{AsqKD*D*-STD}.  Using the numerical estimates in Eqs.~\eqref{AsqKD*D*num},
we obtain the estimate
\begin{eqnarray}
\left|\mathcal{A}\big[K^+  X \pi^- \big]\right|^2 
= |\mathcal{A}[K^0  X \pi^+ ] |^2 &\approx& 
5.7 \times 10^{-10} .
\label{AsqKXpi=}
\end{eqnarray}
We also obtain the upper bounds 
\begin{subequations}
\begin{eqnarray}
\left|\mathcal{A}\big[K^0  X \pi^0 \big]  \right|^2 &<& 
0.17 \times 10^{-10} ,
\label{AsqK0Xpi0<}
\\
\left|\mathcal{A}\big[K^+  X \pi^0 \big]  \right|^2 &<& 
5.0 \times 10^{-10} .
\label{AsqK+Xpi0<}
\end{eqnarray}
\label{AsqKXpi<}%
\end{subequations}

We can use the squared amplitudes in Eq.~\eqref{AsqKXpi=} to estimate branching fractions 
for decays of $B$ into $KX\pi$, with $X \pi$ in the peak from the triangle singularity.
We denote the region of the peak by $(X\pi)_\triangle$.
We declare that region to be $E_{X\pi}$ from 0 up to 
$E_\mathrm{max}= 2 \delta_0= 14.0$~MeV for  $(X\pi^0)_\triangle$
and up to $E_\mathrm{max}= 2 \delta_1= 11.8.0$~MeV  for  $(X\pi^\pm)_\triangle$.
We approximate the integrals over the momentum distributions in 
Eqs.~\eqref{dGammaB+toKpXpi0} and \eqref{dGammaB+toK0Xpi+} using the integral in 
Eq.~\eqref{intsoftpi} and the analogous integral with $\delta_0$ replaced by $\delta_1$.
The resulting estimate of the branching fraction for $B^0 \to K^+  (X \pi^-)_\triangle$
as a function of the binding energy $|E_X|=\gamma_X^2/2\mu$ is
\begin{equation}
\mathrm{Br}\big[B^0 \to K^+  (X \pi^-)_\triangle \big]  \approx
(2.4 \times 10^{-7}) \left( \frac{|E_X|}{0.17~\mathrm{MeV}} \right)^{1/2} \left[ 2.64 - \log \frac{|E_X|}{0.17~\mathrm{MeV}} \right] .
\label{BrKXpi=}
\end{equation}
Our estimate of the branching fraction for $B^+ \to K^0  (X \pi^+)_\triangle$ 
 is larger by the ratio $\Gamma[B^0]/\Gamma[B^+]=1.08$ of the decay widths. 
We  get an upper bound on the branching fraction for $B^0 \to K^0 (X \pi^0)_\triangle$:
\begin{equation}
\mathrm{Br}\big[B^0 \to K^0  (X \pi^0)_\triangle \big]  < 
(8 \times 10^{-8} )\left( \frac{|E_X|}{0.17~\mathrm{MeV}} \right)^{1/2}\left[ 2.82 - \log \frac{|E_X|}{0.17~\mathrm{MeV}} \right] .
\label{BrK0Xpi0}
\end{equation}
Our upper bound on the branching fraction for $B^+ \to K^+ (X \pi^0)_\triangle$
differs only by the replacement of the prefactor by $6 \times 10^{-7}$.

The Belle collaboration has  observed the decay of $B^0$ into $K^+ X \pi^-$  \cite{Bala:2015wep}.
The product of the branching fraction for the $B^0$ decay and the branching fraction for 
the decay $X \to J/\psi\, \pi^+\pi^-$ was measured to be $(7.9\pm1.3\pm 0.4) \times 10^{-6}$. 
Some of the decays come from $B^0 \to K^{*0}X$
followed by the decay of the $K^*(892)$ resonance into $K^+ \pi^-$.
The fraction of events that proceed through the $K^{*0}$  resonance 
is  $(34\pm 9\pm 2)\%$ \cite{Bala:2015wep}. 
Our estimate of the branching fraction for $B^0 \to K^+  (X \pi^-)_\triangle$ in Eq.~\eqref{BrKXpi=} 
implies that the narrow peak from the  charm-meson triangle singularity
can contribute an observable fraction of the decays into $K^+ X \pi^-$
provided the binding energy of the  $X$ is  not too much smaller than 0.17~MeV.


\section{Discussion}
\label{sec:Conclusion}

We have studied the production of $X(3872)$ accompanied by a pion in exclusive decays $B \to K X \pi$.  
This reaction can proceed through the decay of $B$ at short distances into $K$ plus a $D^* \bar D^*$ pair 
with  small relative momentum followed by the rescattering of $D^* \bar D^*$ into $X \pi$.
We used a precise isospin analysis of the decays $B^0 \to K D^{(*)} \bar D^{(*)}$
by Poireau and Zito \cite{Poireau:2011gv} to obtain  approximations for the short-distance amplitudes for these decays.
We verified that those amplitudes are consistent with the measured ratio of the branching fractions 
for $B^+ \to  K^+  X$ and $B^0 \to  K^0 X$, as can be seen in Fig.~\ref{fig:BRp0-eta}.
We used XEFT to calculate the amplitude for the rescattering of $D^* \bar D^*$ into $X \pi$.
The distributions of the kinetic energy $E_{X\pi}$  of $X$ and $\pi$ in the $X \pi$
CM frame are given in Eqs.~\eqref{dGammaB+toKXpi} and \eqref{dBrB0toXpi},
and their shapes are illustrated in Fig.~\ref{fig:Xpi-q}.
The distribution in $E_{X\pi}$ has a narrow peak near the $D^* \bar D^*$ threshold from a 
charm-meson triangle singularity. 
For the decays $B^0 \to K^0 X \pi^0$ and $B^+ \to K^+ X \pi^0$, 
the  peak in $E_{X\pi}$ is predicted to be near $\delta_0= 7.0$~MeV. 
For the decays $B^0 \to K^+ X \pi^-$ and $B^+ \to K^0 X \pi^+$,
the peak is predicted to be near $\delta_1= 5.9$~MeV.

The normalization factors in our $X \pi$ kinetic energy distributions in Eqs.~\eqref{dGammaB+toKXpi} 
and \eqref{dBrB0toXpi} depend on short-distance coefficients in the amplitudes for 
$B \to KD^* \bar D^*$ in Eq.~\eqref{Aij}.  They can be related to coefficients of 
Lorentz-invariant interaction terms constrained by heavy quark-spin symmetry,
such as those in Eq.~\eqref{SSinteraction}.
The simplifying assumption of spin-triplet dominance, which gives the interaction terms in Eq.~\eqref{SSinteraction},
could be eliminated by adding interaction terms for which the $c \bar c$ pair 
is in a spin-singlet state when the charm mesons have equal 4-velocities.
The coefficients of the interaction terms could be determined by squaring the amplitudes 
$\mathcal{A}[B \to K  D^{(*)} D^{(*)}]$, summing over spins, averaging over the Dalitz plot,
and fitting to the results of Ref.~\cite{Poireau:2011gv}.
This would give more reliable estimates of the branching fractions for the decays of $B$ into $KX$ plus a soft pion.
An important limitation of the  isospin analysis of Poireau and Zito is that it assumed that the 
amplitudes were constant across the Dalitz plot.
A more ambitious approach would be to take into account the  variations of the amplitudes 
across the Dalitz plot by fitting the coefficients of the interaction terms to results from
 Dalitz plot analyses of all the decays $B \to K  D^{(*)} D^{(*)}$.  
The BaBar collaboration has carried out  Dalitz plot analyses of the decays $B^0 \to K^+ D^0 D^-$
and $B^+ \to K^+ D^0 \bar D^0$  \cite{Lees:2014abp}.

The region of validity of our expressions for the differential branching fractions
 in Eqs.~\eqref{dGammaB+toKXpi} and \eqref{dBrB0toXpi} is limited to kinetic energy $E_{X\pi}$
 less than about $m_\pi^2/2\mu \approx 10$~MeV.
 The calculations could be extended to larger $E_{X\pi}$ using a strategy applied to
$e^+e^- \to X \gamma$ in Ref.~\cite{Braaten:2019xyz}.  
 After integrating over the loop energy, the amplitudes from the Feynman diagrams in Fig.~\ref{fig:DDtoXpi}
 can be expressed in a form in which the product of the vertex for the coupling of the $X$ to the charm mesons 
 and the  propagators  for those two charm mesons is replaced by the momentum-space wavefunction 
 for the $X$ with momentum $\bm q$.  The results we have presented correspond to the simple wavefunction $\psi(k)$
for $X$ in its rest frame  in Eq.~\eqref{psiX-k}, whose region of validity is limited to $k \ll m_\pi$. 
The wavefunction at $k$ of order $m_\pi$ could presumably be calculated using XEFT.
 Such a wavefunction could be used to extend the calculation of the rate for $B \to K X \pi$ to larger  $E_{X\pi}$.
For  $E_{X\pi}$ larger than about $m_\pi^2/2 \mu_{X\pi} \approx 75$~MeV,
 it is also necessary to use relativistic kinematics for the pion.

We used the assumption of spin-triplet dominance to estimate the branching fractions for decays of $B$
into $K$ plus $X\pi$ in the peak from the charm-meson triangle singularity, which we denoted by $(X\pi)_\triangle$.
Our estimate for $B^0 \to K^+  (X \pi^-)_\triangle$, which applies also to $B^+ \to K^0  (X \pi^+)_\triangle$,
is given in Eq.~\eqref{BrKXpi=}.
We only obtained upper bounds on the branching fractions for $B^0 \to K^0  (X \pi^0)_\triangle$
and $B^+ \to K^0  (X \pi^+)_\triangle$.
These estimates and upper bounds
are essentially proportional to the square root of the binding energy $E_X$ of the $X$.
The Belle experiment at KEK  accumulated roughly $7.7 \times 10^8$ $B \bar B$ events. 
The BaBar experiment at SLAC  accumulated roughly $4.7 \times 10^8$ $B \bar B$ events. 
Our estimates of the branching fractions for $B \to K(X\pi)_\triangle$ suggest that it may   be possible
to observe the narrow peak from the charm-meson triangle singularity in the previous data from those experiments
provided the binding energy of the  $X$ is  not too much smaller than 0.17~MeV.
The prospects are even better at the Belle II experiment at SuperKEKB,
which may  be able to achieve a luminosity 40 times larger than the Belle experiment.
The observation of a peak in the $X\pi$ invariant mass distribution near the $D^* \bar D^*$ threshold
would provide strong support for the identification 
of $X$ as  a weakly bound charm-meson molecule and present a serious challenge to other models.

\begin{acknowledgments}
This work was supported in part by the Department of Energy under grant DE-SC0011726
and by the National Science Foundation under grant  PHY-1607190.
We thank R.~Kass for useful information.
\end{acknowledgments}



\end{document}